\documentclass{aastex631}

\reportnum{2024, ApJ, 977, 236}

\shorttitle{Circumbinary Outflow in W Serpentis}
\shortauthors{Shepard et al.}

\begin{document}

\title{A Spectroscopic and Interferometric Study of W Serpentis Stars. \\ I. Circumbinary Outflow in the Interacting Binary W Serpentis}

\correspondingauthor{Katherine Shepard}
\email{katherine.shepard.1@vanderbilt.edu}

\author[0000-0003-2075-5227]{Katherine Shepard}
\altaffiliation{Physics and Astronomy Department, Vanderbilt University, PMB 401807, 2301 Vanderbilt Place, Nashville, TN 37240, USA}
\affiliation{Center for High Angular Resolution Astronomy and 
Department of Physics and Astronomy, Georgia State University, 
P.O. Box 5060, Atlanta, GA 30302-5060, USA} 

\author[0000-0001-8537-3583]{Douglas R. Gies}
\affiliation{Center for High Angular Resolution Astronomy and 
Department of Physics and Astronomy, Georgia State University, 
P.O. Box 5060, Atlanta, GA 30302-5060, USA} 

\author[0000-0001-5415-9189]{Gail H. Schaefer}
\affiliation{The CHARA Array of Georgia State University, 
Mount Wilson Observatory, Mount Wilson, CA 91023, USA} 

\author[0000-0002-2208-6541]{Narsireddy Anugu}
\affiliation{The CHARA Array of Georgia State University, 
Mount Wilson Observatory, Mount Wilson, CA 91023, USA} 

\author[0000-0002-8376-8941]{Fabien R. Baron}
\affiliation{Center for High Angular Resolution Astronomy and 
Department of Physics and Astronomy, Georgia State University, 
P.O. Box 5060, Atlanta, GA 30302-5060, USA} 

\author[0000-0001-9745-5834]{Cyprien Lanthermann}
\affiliation{The CHARA Array of Georgia State University, 
Mount Wilson Observatory, Mount Wilson, CA 91023, USA} 

\author[0000-0002-3380-3307]{John D. Monnier}
\affiliation{Department of Astronomy, University of Michigan, 
1085 S. University Avenue, Ann Arbor, MI 48109, USA} 

\author[0000-0001-6017-8773]{Stefan Kraus}
\affiliation{Astrophysics Group, School of Physics and Astronomy, 
University of Exeter, Stocker Road, Exeter EX4 4QL, UK} 

\author[0000-0002-0114-7915]{Theo ten Brummelaar}
\affiliation{The CHARA Array of Georgia State University, 
Mount Wilson Observatory, Mount Wilson, CA 91023, USA} 


\begin{abstract}
W~Serpentis is an eclipsing binary system and the prototype of the Serpentid class of variable stars. These are interacting binaries experiencing intense mass transfer and mass loss. However, the identities and properties of both stars in W~Ser remain a mystery. Here we present an observational analysis of high quality, visible-band spectroscopy made with the Apache Point Observatory 3.5 m telescope and ARCES spectrograph plus the first near-IR, long-baseline interferometric observations obtained with the CHARA Array.  We present examples of the appearance and radial velocities of the main spectral components: prominent emission lines, strong shell absorption lines, and weak absorption lines. We show that some of the weak absorption features are associated with the cool mass donor, and we present the first radial velocity curve for the donor star. The donor's 
absorption lines are rotationally broadened, and we derive a ratio of donor to gainer mass of $0.36 \pm 0.09$ based on the assumptions that the donor fills its Roche lobe and rotates synchronously with the orbit. We use a fit of the ASAS light curve to determine the orbital inclination and mass estimates of $2.0 M_\odot$ and $5.7 M_\odot$ for the donor and gainer, respectively. The partially resolved interferometric measurements of orbital motion are consistent with our derived orbital properties and the distance from Gaia EDR3. Spectroscopic evidence indicates that the gainer is enshrouded in an opaque disk that channels the mass transfer stream into an outflow through the L3 region and into a circumbinary disk.  
\end{abstract}

\keywords{}

\section{Introduction} 
\label{sec:introduction}

Massive stars are often found in close pairs that are destined to become interacting binaries \citep{deMink2013}. The mass transfer and loss that occurs during their interaction will dramatically alter their evolutionary paths and change the properties of collapsed remnants. As a massive star evolves, its primary source of fuel will switch from H core fusion during the main sequence phase to H-shell fusion during the red giant phase. This results in the outer atmosphere of the star expanding until it fills the Roche lobe. 
This can result in Roche lobe overflow and mass transfer \citep{Paczynski1971,Thomas1977,vanRensbergen2011,Tauris2023}. As gas moves from the larger star (mass donor) to the smaller star (mass gainer), the separation of the stars and the orbital period will decrease. This continues until the mass ratio reverses and the mass donor is now less massive than the mass gainer. When this occurs, the orbital period and semimajor axis begin to increase as mass transfer continues,
provided that the process is conservative, i.e., no systemic mass loss occurs.

This period of rapid evolution is a defining characteristic of the W Serpentis class (also known as Serpentid binary systems; \citealt{Plavec1980}). Some of these systems will experience non-conservative mass transfer, i.e., a portion of the mass transferred is ejected from the inner binary into a circumbinary disk (CBD). The gas stream from the mass donor will form an accretion disk around the mass gainer. The size of this disk is limited by the Roche lobe of the mass gainer. Once the gainer is spun up to near critical rotation, gas accretion is hindered, and the gas stream will fill an enlarged, circumstellar torus that may block the star from view. If mass transfer continues to occur, the additional material will be expelled from the disk. This material will gradually travel outwards and may begin to condense, forming a CBD \citep{Nazarenko2005}. Understanding what happens to this expelled material and the fraction of mass lost by these systems is crucial to our models of stellar evolution. \citet{Lu2023} constructed models of systems experiencing mass transfer and calculated the fraction of mass lost to the circumbinary disk. They find that a mass-transfer rate exceeding approximately $10^{-4}$ $M_{\odot}$~yr$^{-1}$ results in an ``order-unity fraction'' of transferred mass that is lost through the L3 Lagrangian point. The fraction of mass lost depends on gas metallicity, radiative cooling rate, viscosity of the gas, and the binary mass ratio. 

W~Serpentis (e.g., HD 166126, BD-15$^\circ$4842) is the prototype of the Serpentid binary stars. It is an eclipsing binary with an orbital period of $14.17$ days \citep{Gaposchkin1937,Bauer1945, Guinan1989,Erdem2014}. An outline of the long observational history of W~Ser is given by \citet{Guinan1989} and \citet{Weiland1995}. Their primary conclusions are summarized here. The light curve displays only one eclipse and is marked by large stochastic variations \citep{Guinan1989}. The visible spectrum is complex and has six distinct absorption and emission line kinematical components \citep{Sahade1957,Barba1993} including disk-related emission features, forbidden line species, and sharp, stationary shell lines. The deep absorption line, shell component has the appearance of an F8/G2~Iaep star \citep{Houk1988}, and the continuum shape in the visible regime is consistent with a temperature associated with this classification. Spectroscopic investigations by \citet{Sahade1957} and \citet{Hack1958} show that the hotter component is the object that is eclipsed at primary minimum (phase $0.0$).
The plethora of emission lines in the visible spectrum indicate the presence of circumstellar and circumbinary material related to active mass transfer and loss. One primary distinguishing feature of W~Ser and other Serpentids is the presence of UV emission lines \citep{Wilson1989} such as \ion{N}{5}, \ion{Si}{4}, \ion{C}{4}, \ion{C}{2}, \ion{O}{3}, \ion{Si}{3}, and \ion{He}{2} \citep{Plavec1981,Plavec1982a,Plavec1982b,Plavec1989,Sanad2013}. The UV emission lines appear to vary in strength with orbital phase \citep{Sanad2013}. \citet{Weiland1995} analyzed the profiles of the \ion{Si}{4} $\lambda 1400$ doublet using spectra from the Hubble Space Telescope Goddard High Resolution Spectrograph, and they developed a model that relates the emission line flux to mass transfer. They and \citet{Deschamps2015} note the presence of an elevated continuum flux in the far-ultraviolet (FUV), one that exceeds the level expected for an F-supergiant. This suggests that there is another hot, hidden flux component in the system. 

The mass loss models created by \citet{Lu2023} are likely applicable to the stage that W~Ser is currently experiencing (see their Fig.~$1$ and Fig.~$6$). Several studies have suggested that the F-supergiant continuum flux actually forms in the pseudo-photosphere of an accretion disk that surrounds and obscures the mass gainer \citep{Guinan1989,Plavec1989,Weiland1995,Erdem2014}. The emission lines can form in both the hot plasma found in the inner portion of the accretion disk and at the point of contact between the mass transfer stream and the disk. The latter ``hot spot'' may contribute to the flux at short wavelengths \citep{Weiland1995}. The circumbinary gas is responsible for narrow and stationary emission lines. The spectral signatures of both the mass gainer and mass donor stars remain undetected. There is considerable uncertainty surrounding the orbital properties, and estimates of the total mass range from $2.5 M_\odot$ \citep{Mennickent2016} to $7.1 M_\odot$ \citep{Erdem2014}. The associated semimajor axis estimates range from $0.16$ to $0.22$ AU. The estimated distance from Gaia EDR3 is $857_{-15}^{+14}$ pc \citep{Bailer-Jones2021}, so the angular semimajor axis is estimated to be in the range of $0.18$ to $0.26$ mas. This is smaller than the nominal angular resolution of long baseline interferometers like the CHARA Array.
Nevertheless, although the binary may not be fully resolved, a larger circumbinary disk might be resolvable with interferometry.

This paper is the first in a series that will investigate a sample of nine systems that are probable Serpentids. We will analyze their optical spectra with the aim of obtaining radial velocity curves, and updated orbital periods and elements. We will also present interferometric data with the aim of estimating the extent of the CBDs, and creating simple models for each system. This paper focuses on W~Ser itself and is structured as follows. In Section \ref{sec:contemporary-ephemeris} we calculate an updated contemporary ephemeris and orbital period. In Section \ref{sec:spectroscopy} we delve into the details of the spectroscopy. We discuss the various features that can be found in the spectrum of W~Ser, including the shell lines (Section \ref{sec:shell-lines}), emission lines (Section \ref{sec:emission-lines}), features from the mass donor (Section \ref{sec:donor-star-photospheric-lines}), and lines from the vicinity of the mass gainer (Section \ref{sec:doppler-tomography-gainer}). We discuss their radial velocities and probable origins. In Section \ref{sec:chara-interferometry} we describe our interferometric observations made with the CHARA Array, and the models we made using these data. Finally, we discuss our findings (Section \ref{sec:discussion}) and draw several novel conclusions (Section \ref{sec:conclusions}) regarding this mysterious system.

\newpage
\section{Contemporary Ephemeris}
\label{sec:contemporary-ephemeris}
\begin{figure*} 
        \centering
	\includegraphics[width=17cm]{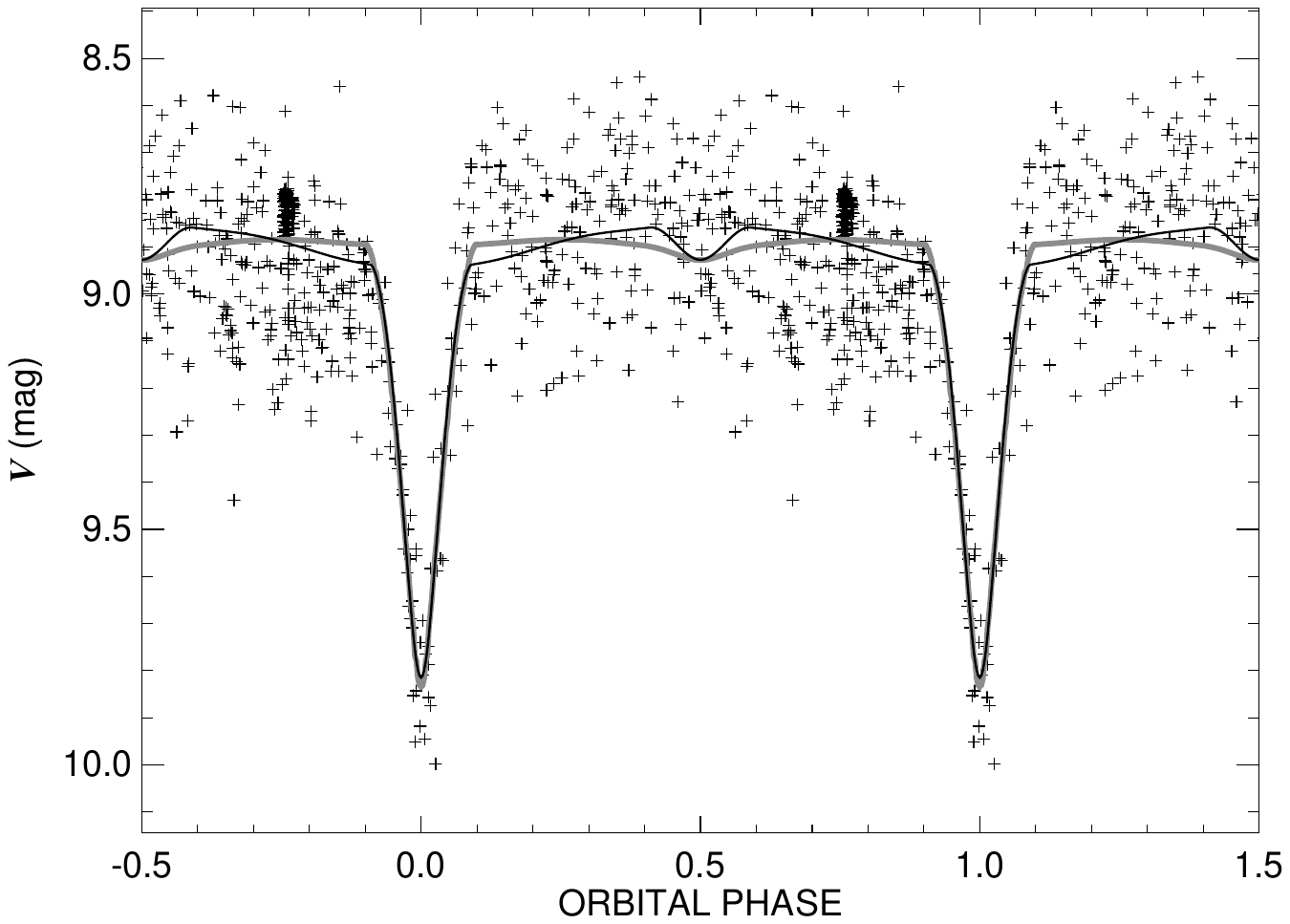}
	\caption{The $V$-band light curve from the ASAS Catalogue of Variable Stars with two ELC fits over-plotted. The solid line is a fit made by assuming that the gainer has a typical B-type star temperature $T({\rm gainer})=17900$ K but large size ($R_g = 13.2~R_\odot$). The thick gray line is a second fit with a small gainer ($R_g = 3.8~R_{\odot}$) surrounded by a large, optically thick disk ($R_d = 14.3~R_{\odot}$).}
        \label{fig:asas-light-curve-elc-fit}
\end{figure*}

We need an accurate orbital ephemeris in order to study the spectroscopic and interferometric variations with orbital phase. The timing of the middle of the primary eclipse would ideally come from a contemporary light curve. The most recent light curve is unfortunately over a decade old. This $V$-band light curve is from the All-Sky Automated Survey (ASAS) Catalogue of Variable Stars\footnote{http://www.astrouw.edu.pl/asas/?page=acvs} \citep{Pojmanski2004} and is shown in Figure \ref{fig:asas-light-curve-elc-fit}. This conundrum is further exacerbated by the fact that W~Ser's orbital period is increasing. \citet{Erdem2014} collected published times of mid-primary-eclipse for W~Ser that span the past century and compared the data to the predicted eclipse times given the current estimated orbital period. They found that the Observed minus Calculated (O-C) times of primary eclipse showed a general rate of period increase of $18.8 \pm 0.4$ s~yr$^{-1}$. 
This is one of the largest period increase rates found by \citet{Erdem2014} among similar systems, and 
it is consistent with expectations for mass transfer following the mass ratio reversal for the conservative case (see Section 5.2).
Due to this period increase and the lack of an updated orbital period, we need to calculate the current orbital period given the period from ASAS from $2006$ and the period increase since then. 

The epoch of primary eclipse reported by the ASAS Catalogue is ${T (ASAS)} = {\rm HJD}~2,451,950.3$ and the associated orbital period is $P{(ASAS)} = 14.174658~{\rm days}$. First we checked the reported epoch by fitting an Eclipsing Light Curve (ELC) model \citep{Orosz2000} to the ASAS $V$-band photometry. This fit, and an additional fit we tested (discussed in Section \ref{sec:discussion-prelim-light-curve}), are over-plotted on the ASAS data in Figure \ref{fig:asas-light-curve-elc-fit}. Using the ASAS period and the light curve fit, we find a revised epoch of the primary eclipse for a time near the middle for the ASAS observations of $T{(MID)} = {\rm HJD}~2,453,268.7632$ (or $0.22$ days later than the ASAS epoch). 

We then extrapolated forward in time to the approximate epoch of our observations. This corresponded to $475$ orbits after the central ASAS epoch or $18.4$ years later. We calculated our updated period as follows: 
\begin{equation} \label{eq:1}
    P(NEW) = P(ASAS) + \dot{P} \triangle t = 14.17867 ~{\rm days.}
\end{equation}
Next we calculated the predicted O-C difference after this elapsed time:
\begin{equation} \label{eq:2}
    O-C = 0.5 P(ASAS) \dot{P} \epsilon^2 = 0.9526 ~{\rm days}
\end{equation}
where $\epsilon = 475$ elapsed orbital cycles \citep{Kalimeris1994}. Finally, we calculated the contemporary epoch,
\begin{equation} \label{eq:3}
    T(NEW) = T(MID) + \epsilon P(ASAS) + (O-C) = {\rm HJD}~2,460,002.6783.
\end{equation}
We will use $P(NEW)$ and $T(NEW)$ to determine orbital phase throughout this work. We define phase $0.0$ as the time of primary eclipse when the hotter object is eclipsed by the cooler object.

\section{Spectroscopy}
\label{sec:spectroscopy}
We obtained $10$ optical spectra of W~Ser between $2021$ and $2023$. These were made with the Astrophysical Research Consortium Echelle Spectrograph (ARCES) mounted on the $3.5$ m telescope at Apache Point Observatory (APO). These are high-resolution (R$\approx 31500$) optical spectra \citep{Wang2003} covering $3200$-$10000$ \AA ~over $107$ orders. Our exposure times were $20$ minutes on average, yielding a S/N $=200$ in the better exposed portions of the spectrum. Heliocentric Julian dates at mid-exposure are listed in Table \ref{table:rad-vels}. The wavelength calibration was set by referencing the Thorium-Argon (ThAr) spectrum taken closest to the time of observation. Spectra were reduced using the reduction guide written by Julie Thorburn and the reduction cookbook by Karen Kinemuchi\footnote{Both can be found on the APO Wiki: http://astronomy.nmsu.edu/apo-wiki/doku.php}. These guides describe standard IRAF procedures, including bias subtraction, flat-fielding, scattered light subtraction, spectrum extraction, and wavelength calibration. We divided out the echelle blaze function present in the shape of the continuum for each order by making polynomial fits of the orders that are free of strong absorption or emission features. We then used these fits to interpolate across those orders that contained strong and broad lines, such as the H-Balmer series \citep{Kolbas2015}. The flux-normalized spectra were then transformed to the heliocentric frame, and the individual echelle orders were merged onto a standard logarithmic wavelength grid. Finally, the telluric features in the vicinity of H$\alpha$ were removed.

The spectrum of W~Ser is complex making detection of spectral features originating from either stellar component very difficult. Our aim is to determine the origin of the main spectral components. These features include very deep and narrow ``shell'' lines (found in metal transitions such as \ion{Fe}{1} and \ion{Ti}{2} and the cores of the H-Balmer lines), H-Balmer emission lines, and forbidden transitions \citep{Bauer1945,Sahade1957,Hack1958,Barba1993}. In the following we present examples of the shell lines (Section \ref{sec:shell-lines}), emission lines (Section \ref{sec:emission-lines}), photospheric lines associated with the mass donor (Section \ref{sec:donor-star-photospheric-lines}), and features that track the predicted motion of the mass gainer and its circumstellar disk (Section \ref{sec:doppler-tomography-gainer}).

\newpage

\setcounter{table}{0}
\setlength\LTleft{0pt}
\setlength\LTright{0pt}
\begin{scriptsize}
\begin{longtable}{@{\extracolsep{\fill}}cccrc@{}}
\caption{\\ Radial Velocities for W Serpentis}\\
\toprule
\multicolumn{1}{c}{Date} & \multicolumn{1}{c}{Orbital}  & \multicolumn{1}{c}{$V_r$ (Shell)} &  \multicolumn{1}{c}{$V_r$ (Donor)}  & \multicolumn{1}{c}{Residuals (O-C)} 
\\
\multicolumn{1}{c}{(HJD - $2,400,000$)} & \multicolumn{1}{c}{Phase}  & \multicolumn{1}{c}{(km s$^{-1}$)} &  \multicolumn{1}{c}{(km s$^{-1}$)}  & \multicolumn{1}{c}{(km s$^{-1}$)} 
\endhead
\hline
\endfoot 
\hline
59336.9653 & 0.0483 & $-27.0 \pm 0.7$ & $14.1 \pm 3.4$     & \phn\phs$3.8$     \\ 
59358.8160 & 0.5894 & $-18.0 \pm 1.3$ & $-102.3 \pm 3.1$   & \phn$-8.1$    \\ 
59364.8298 & 0.0135 & $-27.2 \pm 0.8$ & $-14.9 \pm 2.9$    & \phn\phs$1.7$     \\ 
59409.6750 & 0.1764 & $-27.9 \pm 0.8$ & $91.9 \pm 5.2$     & \phn\phs$6.6$     \\ 
59741.8275 & 0.6026 & $-19.0 \pm 1.0$ & $-112.5 \pm 3.0$   & \phn$-9.7$    \\ 
60098.8330 & 0.7816 & $-22.3 \pm 1.0$ & $-129.6 \pm 5.8$   & \phs$20.9$    \\ 
60106.8775 & 0.3490 & $-32.3 \pm 1.0$ & $84.6 \pm 3.8$     & \phn\phs$9.7$     \\ 
60108.9064 & 0.4921 & $-25.0 \pm 1.1$ & $-11.8 \pm 5.2$    & \phn\phs$9.2$     \\ 
60202.5941 & 0.0998 & $-29.6 \pm 0.9$ & $26.4 \pm 3.7$     & $-20.1$   \\ 
60223.5706 & 0.5792 & $-22.0 \pm 1.2$ & $-101.2 \pm 3.2$   & $-13.9$    
\label{table:rad-vels}
\end{longtable}
\end{scriptsize}

\subsection{Shell Lines}
\label{sec:shell-lines}
\begin{figure} 
    \centering
       \includegraphics[width=0.4\linewidth]{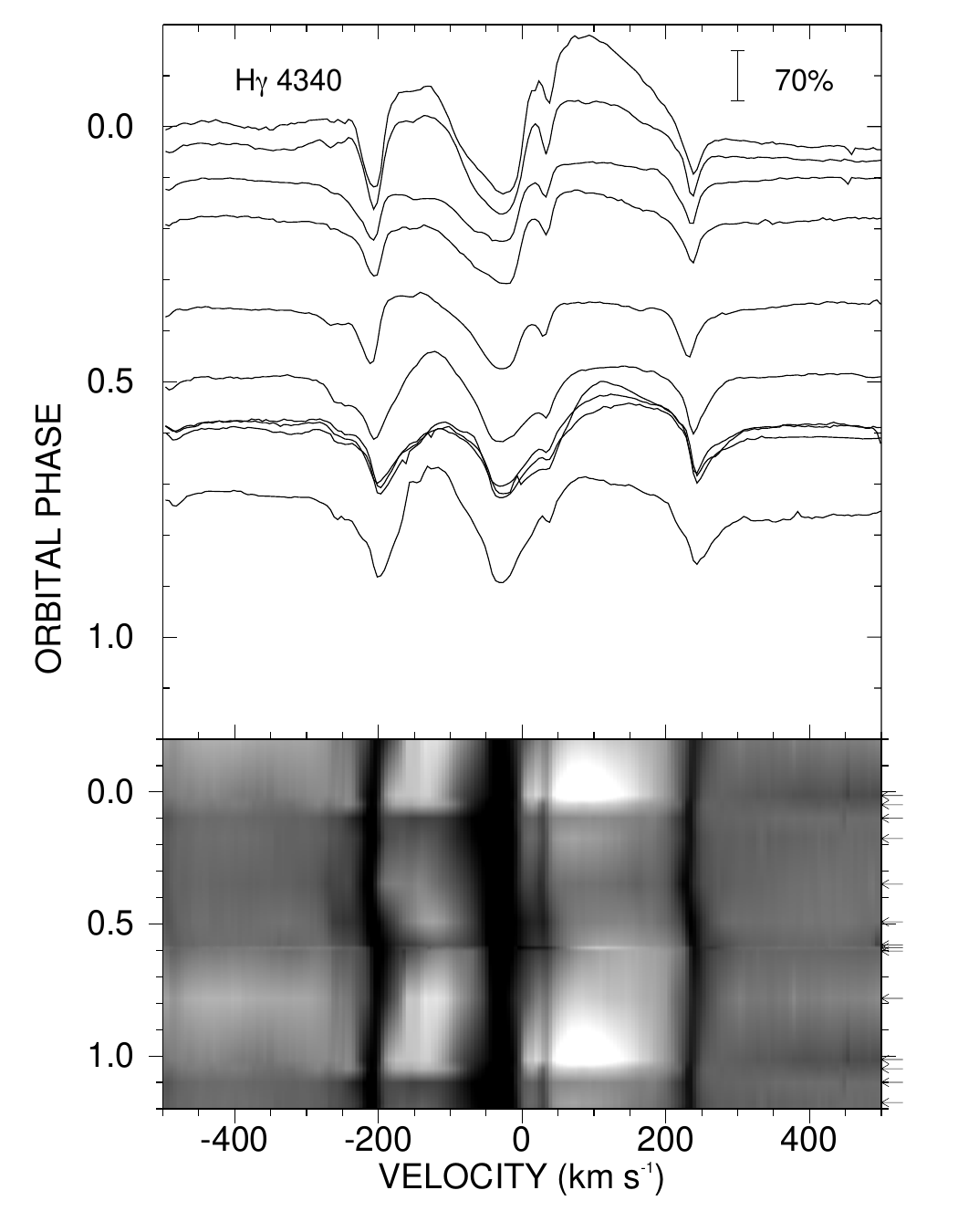}
       \includegraphics[width=0.4\linewidth]{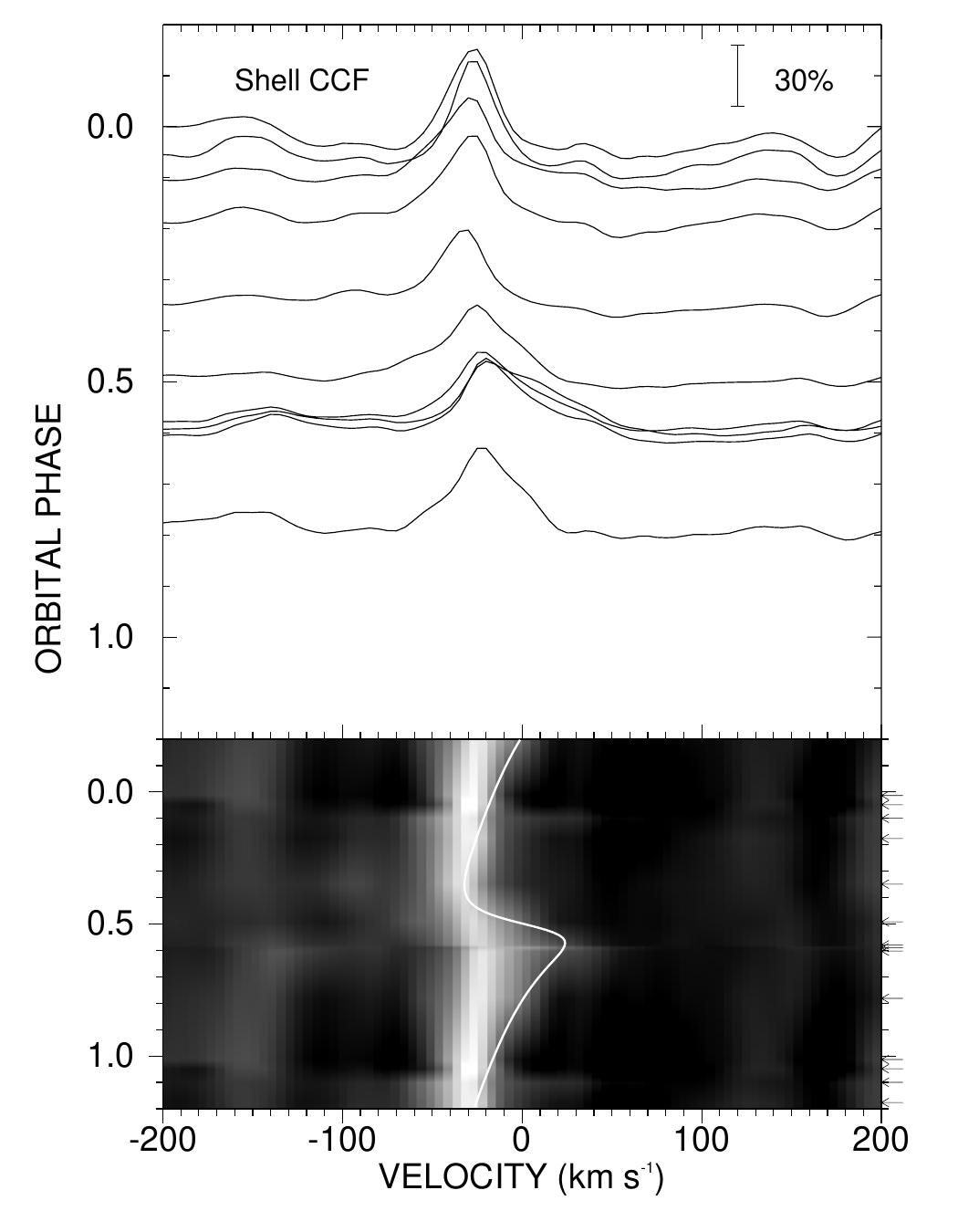}
    \caption{\textbf{Left -} The top panel shows each spectrum as a function of radial velocity for H$\gamma$ $\lambda 4340$ with the normalized continuum offset to the associated orbital phase of the observation. The bottom panel shows the same plotted as a grayscale intensity between deepest absorption (black) and brightest emission (white) interpolated across orbital phase. H$\gamma$ appears in the center and is flanked by shell lines of \ion{Ti}{2} $\lambda 4337.925$ ($-201$ km~s$^{-1}$), \ion{Ti}{2} $\lambda 4341.371$ ($+37$ km~s$^{-1}$), and \ion{Ti}{2} $\lambda 4344.290$ ($+239$ km~s$^{-1}$), among other weaker lines. \textbf{Right -} Cross-correlation functions of the shell lines with a model template spectrum.  These CCFs are offset vertically with the local background shifted to the associated orbital phase of the observation. The lower panel is a grayscale representation of the CCFs with the orbital velocity curve from \citet{Bauer1945} overplotted as a solid white line.}
    \label{fig:hgamma-shellCCF-greyscales}
\end{figure}

Sharp, deep absorption lines dominate the continuum throughout the spectrum of W~Ser. These lines mainly consist of transitions of \ion{Fe}{1}, \ion{Fe}{2}, and \ion{Ti}{2}. The strongest features are listed by \citet{Bauer1945} (see his Table 2). Several of these shell lines are close neighbors to H$\gamma$ $\lambda 4340$, such as \ion{Ti}{2} $\lambda\lambda 4337, 4341, 4344$. These features are visible in the left plot of Figure \ref{fig:hgamma-shellCCF-greyscales}. They appear consistently blue-shifted at a velocity of approximately $-25$ km~s$^{-1}$.

Measurements of shell line velocities were used in earlier work as possible indicators of orbital motion. Many of the features are blends, so we are not able to measure individual lines easily. Instead we measured the shell line velocities by cross-correlating a model with a region of the spectrum. We selected a wavelength range of $4550$ - $4600$ \AA ~(see Fig.\ \ref{fig:plottomabs} below) because it lacked emission lines and the unblended, moderate-strength shell features all showed similar phase-related behavior. We utilized a model spectrum from the Pollux Synthetic Stellar Spectra Database\footnote{https://pollux.oreme.org}. We selected an AMBRE/MARCS model \citep{Gustafsson2008} with $T_{\rm eff} = 6750$~K, $\log g = 1$, microturbulence $=1$ km~s$^{-1}$, and solar abundances. We rebinned this model onto our observed wavelength grid. The resulting cross-correlation functions (CCFs) are shown on the right side of Figure \ref{fig:hgamma-shellCCF-greyscales}. We determined the radial velocity of the shell lines by fitting a parabola to the peak of the CCFs. These radial velocities are listed in column 3 of Table \ref{table:rad-vels} under the heading $V_r$(Shell).

The shell line CCFs show modest radial velocity variations that are unrelated to the orbital motion of the stellar components. First, the variation is much smaller than expected for binary motion. Second, the largest variation occurs immediately after phase $\phi=0.5$, instead of near the expected extrema at $\phi=0.25$ and $\phi=0.75$ for a circular orbit. Third, if we interpret this motion as Keplerian motion in an elliptical orbit, then the implied time of conjunction is inconsistent with the light curve ephemeris, as discussed by \citet{Bauer1945}. 

The shell line velocities were previously measured by \citet{Bauer1945}, and their maximum amplitude has changed significantly since his observations made in 1943. The bottom portion of the right panel in Figure \ref{fig:hgamma-shellCCF-greyscales} illustrates the motion we see as a thick and almost stationary band.  The historical orbital solution from \citet{Bauer1945} is over-plotted as a solid white line that has a much larger range. 
Our spectra were obtained about $80$ years later, 
and the shell line velocities now have a maximum radial velocity that is about $40$ km~s$^{-1}$ smaller than found by Bauer at the same phase. There is only a weak excess in the red wing of the CCF that reaches a velocity comparable to that observed by Bauer. 
Bauer considered a possible elliptical orbital fit of the velocities, but the resulting small amplitude and phase of periastron were inconsistent with the orbital light curve. 
He concluded that the motion of the shell lines is not connected to the orbital motion of the binary. 

We suggest that the shell lines are not formed within the inner binary, but form instead in a large CBD that is centered on the binary's center of mass and that partially obscures the stellar flux. The CBD gas motions are mainly orthogonal to the line of sight towards the binary, so no orbital motion is observed.

\subsection{Emission Lines}
\label{sec:emission-lines}
\begin{figure*} 
        \centering
	\includegraphics[width=0.4\linewidth]{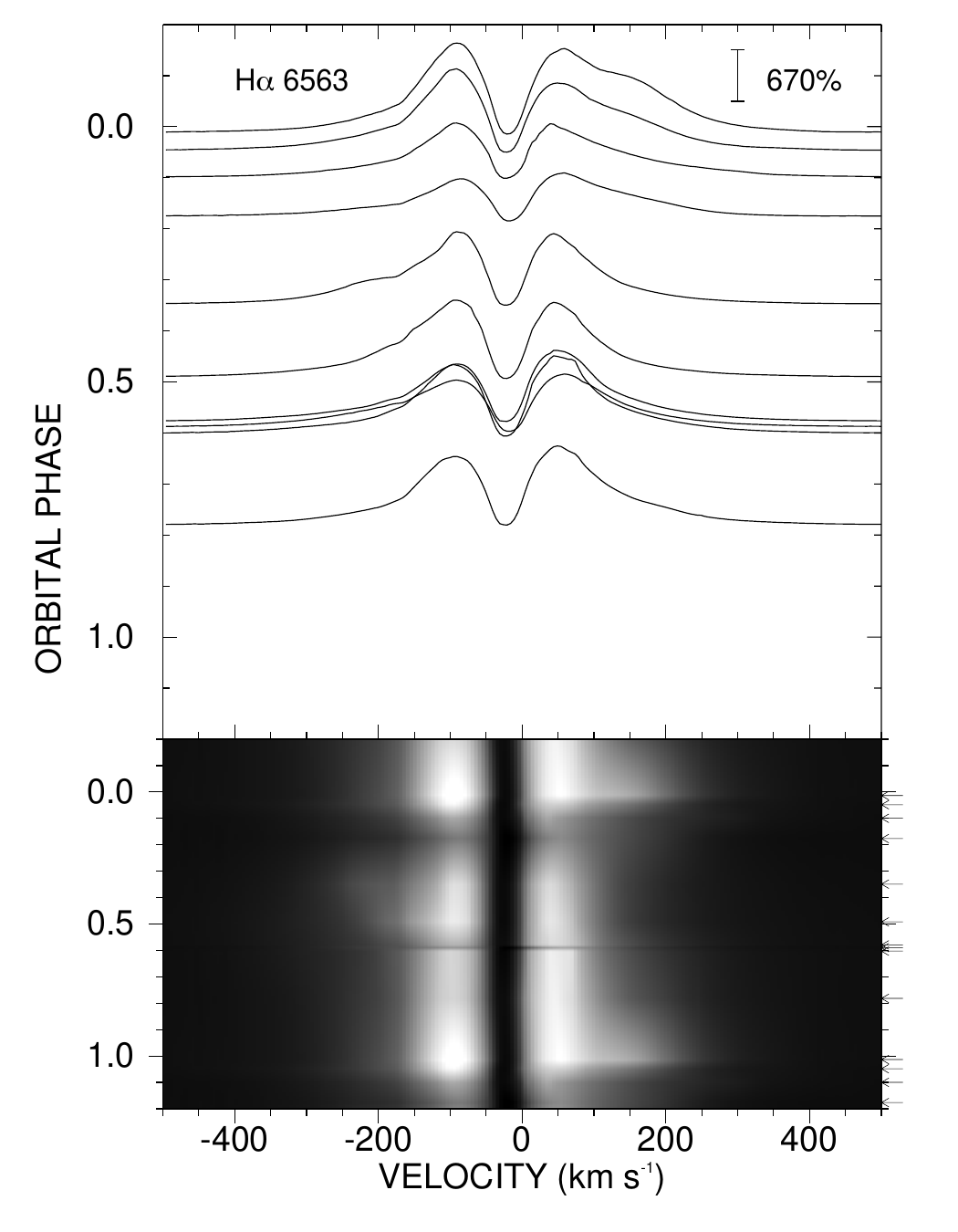}
 	\includegraphics[width=0.4\linewidth]{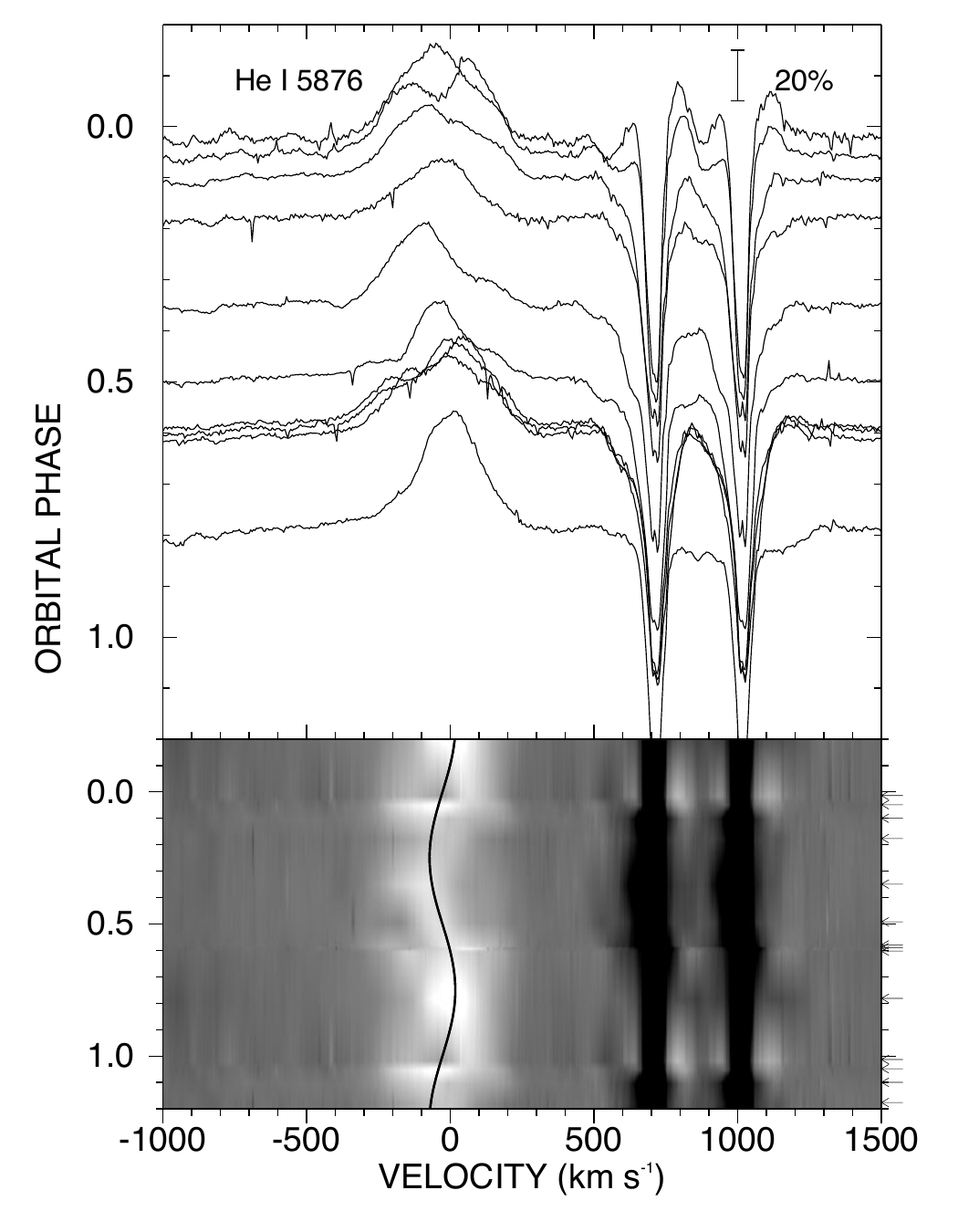}
	\caption{\textbf{Left -} H$\alpha$ $\lambda 6563$ profiles (top panel) and the associated grayscale diagram (bottom panel) in the same format as Fig.~2. The normalized continuum is shifted to the orbital phase of observation. \textbf{Right -} \ion{He}{1} $\lambda 5876$ profiles. Our estimated radial velocity curve for the mass gainer is over-plotted on the grayscale diagram as a solid black line.
        }
        \label{fig:greyscales-halpha-HeI5876}
\end{figure*}

\begin{figure*} 
        \centering
	\includegraphics[width=0.4\linewidth]{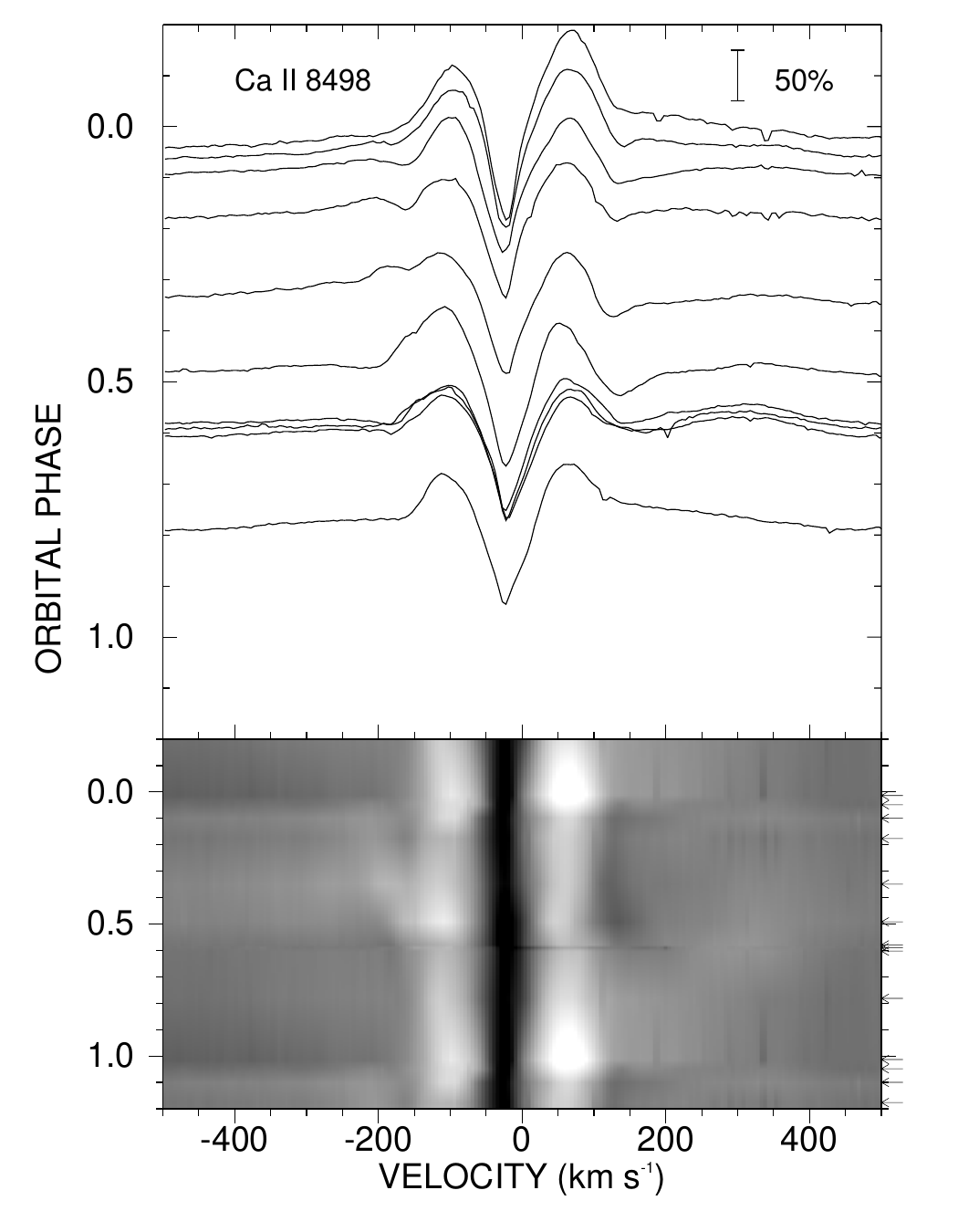}
	\includegraphics[width=0.4\linewidth]{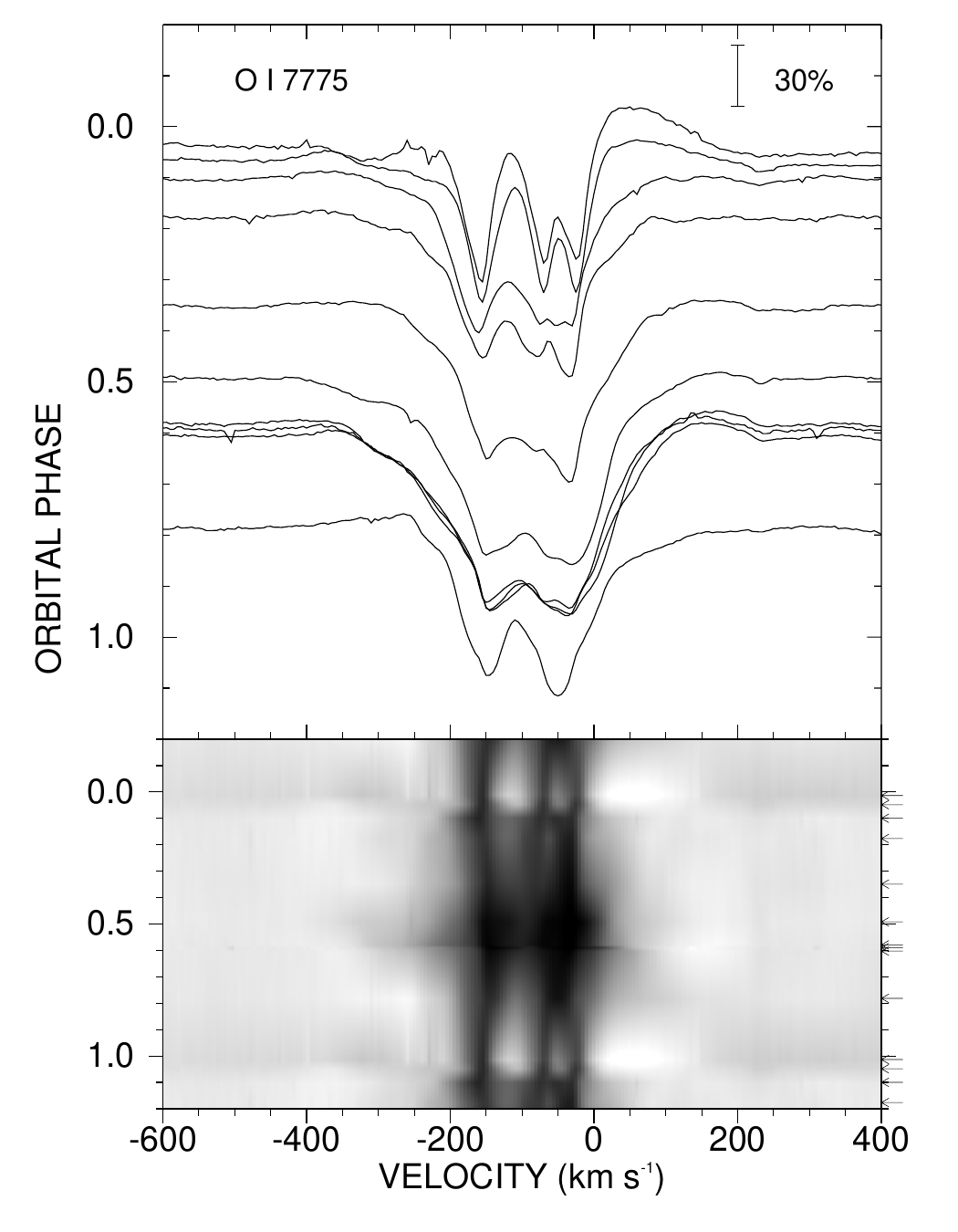} 
	\caption{\textbf{Left -} \ion{Ca}{2} $\lambda 8498$ profiles in the same format as Fig.~2. \textbf{Right -} \ion{O}{1} $\lambda 7775$ profiles. 
        }
        \label{fig:greyscales-CaII8498-OI7775}
\end{figure*}

The second kind of features that dominate the spectra of W~Ser are strong emission lines. 
Emission features may form in multiple locations such as the gas stream between donor and gainer, the torus surrounding the 
gainer, gas leaving the binary, and gas in the circumbinary disk. 
Each location is associated with a specific pattern of orbital Doppler shifts, and here we use the observed variations to infer the dominant emission source.
The emission lines in the visible-band include H$\beta$ $\lambda 4861$, \ion{He}{1} $\lambda 5875$, H$\alpha$ $\lambda 6563$, \ion{Ca}{2} $\lambda 8498$, and P11 $\lambda 8862$ (and other Paschen H lines). Figure \ref{fig:greyscales-halpha-HeI5876} and Figure \ref{fig:greyscales-CaII8498-OI7775} show a compilation of the orbital variation diagrams for some of these lines. 
The H$\alpha$ and \ion{Ca}{2} lines appear to be double peaked, indicating the presence of a large disk.  However, neither H$\alpha$ nor \ion{Ca}{2} show radial velocity variations similar to what we would expect for binary motion. Therefore, we conclude that these lines are formed in the CBD. However, we do see distinct changes in the lines that appear to depend on phase. For example, the plot of H$\alpha$ shows an extended red wing at $\phi=0.0$ and a slight excess in the blue wing around $\phi=0.5$ (discussed below in Section \ref{sec:discussion}). 

The grayscale plot for \ion{He}{1} $\lambda 5876$ reveals complex variations in strength and shape. The spectral feature is occasionally double peaked (around $\phi=0.1$ and $\phi=0.6$) but most of the time it only has a single peak. 
In the next section, we present a radial velocity curve for the mass donor, which together with the mass ratio leads to an estimated radial velocity curve for the mass gainer. This predicted radial velocity curve for the mass gainer is over-plotted on the grayscale panel for \ion{He}{1} $\lambda 5876$ as a solid black line. We suggest that the similarities between the predicted radial velocity motion of the mass gainer and the observed radial velocity motion of the \ion{He}{1} line indicate that the emission originates in the gas surrounding the mass gainer. We expect that the mass gainer has reached a mass and temperature similar to that of a B-star, and its flux would heat the surrounding gas. Because this emission line is associated with hot gas, we can assume that it does not form out in the CBD where the gas has a cooler temperature than gas closer to the stars. This supports the idea that the \ion{He}{1} line originates in the heated area surrounding the hot mass gainer.

We caution the reader on the appearance of enhanced emission lines at $\phi=0.0$. This is not physical, but is a result of the flux normalization process. During the eclipse, the brighter gainer is eclipsed by the fainter donor, and the overall continuum flux from the system decreases. By normalizing the continuum to unity, we artificially enhance any other uneclipsed flux sources. This results in stronger emission profiles as seen in Figure \ref{fig:greyscales-halpha-HeI5876} and Figure \ref{fig:greyscales-CaII8498-OI7775} around $\phi=0.0$. This enhanced emission at the eclipse phase is also seen in H$\gamma$ $\lambda 4340$ (left panel of Fig.~\ref{fig:hgamma-shellCCF-greyscales}).

\subsection{Donor Star Photospheric Lines}
\label{sec:donor-star-photospheric-lines}
The mass transfer rate will peak when the stars attain their minimum separation and the mass ratio reverses. By this stage, the donor star will have a cooler surface temperature (see the predicted HR-diagram track in Fig.~2 of \citealt{Gotberg2018}). This large difference in temperature between the components is probably the reason for the contrast between the deep primary eclipse and the hidden secondary eclipse (see Fig.~\ref{fig:asas-light-curve-elc-fit}).

Based upon the predictions for the expected temperature of the mass donor, we selected a model spectrum for the donor from the  AMBRE/MARCS grid from the Pollux Synthetic Stellar Spectra Database.
The parameters of the model for the donor are $T_{\rm eff} = 5000$ K, $\log g=2.5$, microturbulence $=1$ km~s$^{-1}$, and solar abundances. The model was then rebinned onto the observed wavelength grid. We then compared the model to the observed spectrum obtained during primary eclipse when the donor contributes relatively more light.

\begin{figure*} 
        \centering
	\includegraphics[angle=0,width=15cm]{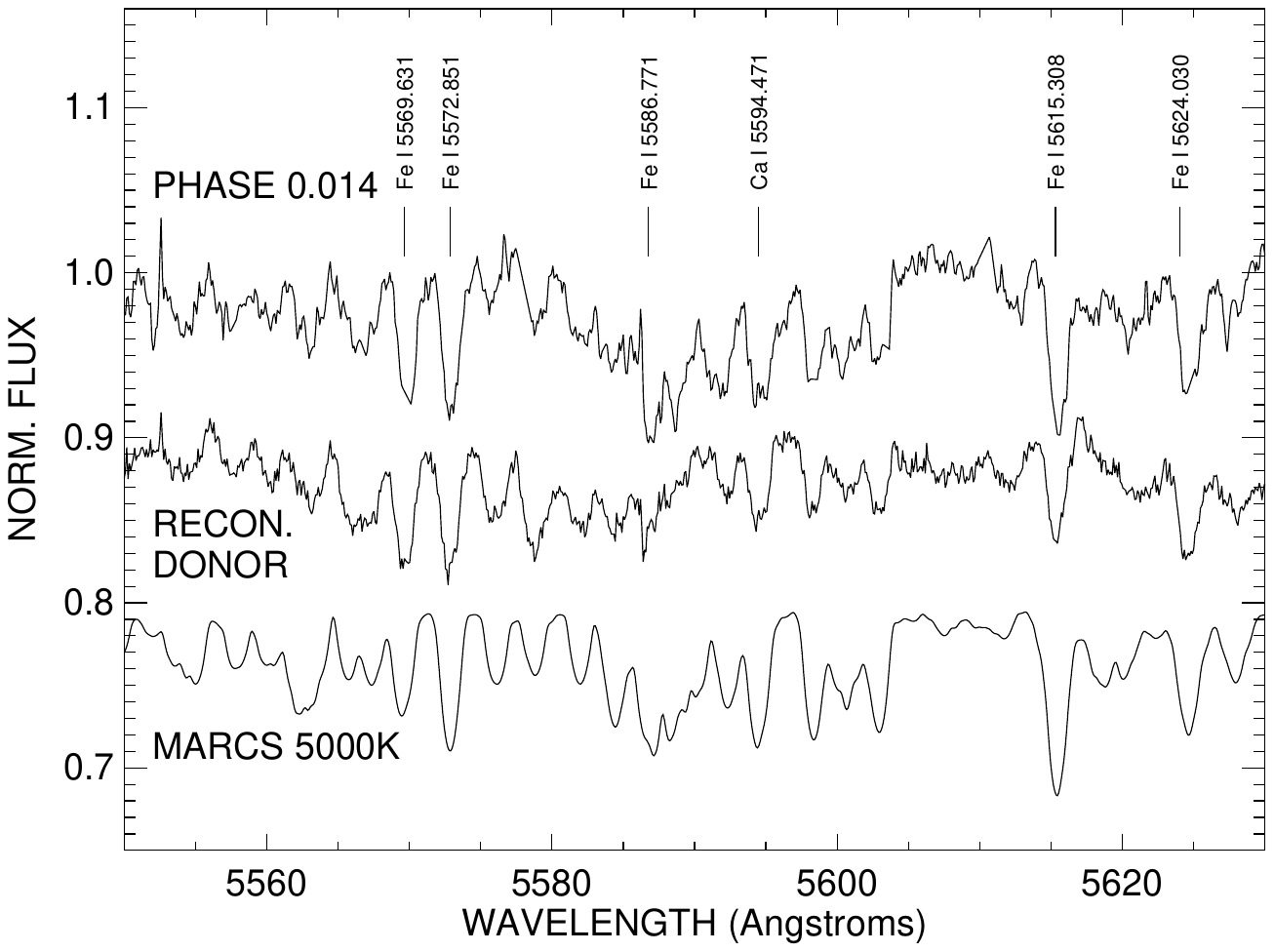}
	\caption{\textbf{Top -} Observed APO/ARCES spectrum of W~Ser obtained near the primary eclipse (HJD 2,459,364.830, $\phi = 0.014$) in the rest frame with identifications of several prominent lines. \textbf{Middle -} Tomographic reconstruction of the donor's spectrum offset for clarity. \textbf{Bottom -} MARCS model spectrum for $T_{\rm eff} = 5000$ K, rebinned to the observed wavelength, broadened to the apparent projected rotational velocity of the donor star, rescaled for additional continuum flux from the system, and offset for clarity. 
        }
        \label{fig:donor-model-lines}
\end{figure*}

Locating line features similar to those seen in the model proved challenging. This is due to the overwhelming and consistent presence of the shell lines, the donor contributing less light at shorter wavelengths, and the presence of emission and telluric lines at longer wavelengths. We did find that the $5550-5630$ \AA ~range is relatively free of shell lines, contains no emission or telluric lines, and appears to show several prominent stellar lines (mostly \ion{Fe}{1}). There is a strong resemblance between the observed and model spectra in this region as shown in Figure~\ref{fig:donor-model-lines}.  

\begin{figure*} 
        \centering
	\includegraphics[angle=0,width=15cm]{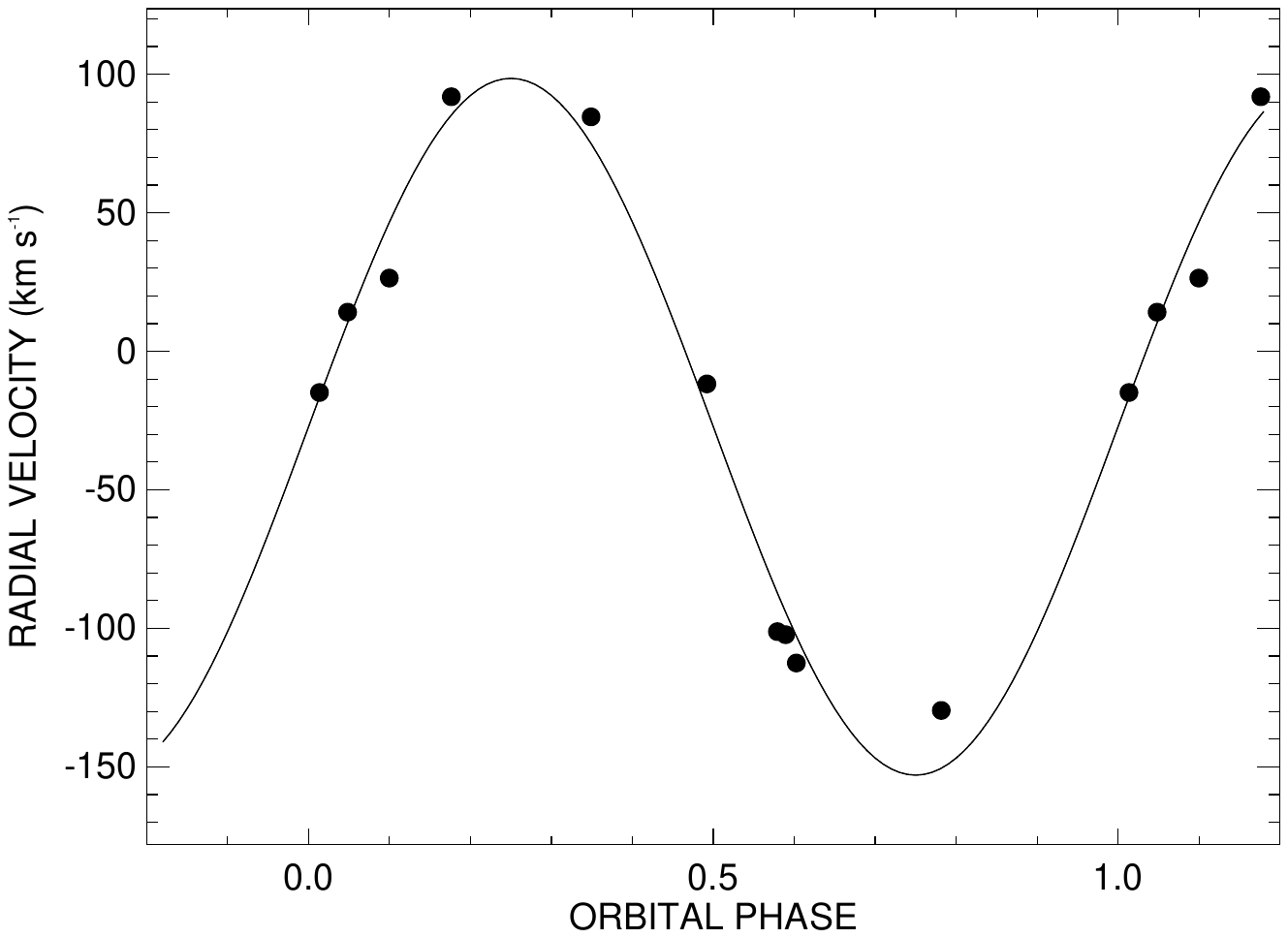}
	\caption{CCF radial velocity measurements (points) for the donor's spectral lines in the region seen in Fig.\ \ref{fig:donor-model-lines}, over-plotted with a circular orbital fit (solid line).
        }
        \label{fig:donor-rv-curve}
\end{figure*}

We expect the spectrum of the mass donor to be present, although weaker, at other orbital phases. Therefore, we calculated the CCFs of the observed and model spectra for the region plotted in Figure~\ref{fig:donor-model-lines} to determine radial velocities for the donor star. We succeeded in finding a measurable peak for each of our observations. We then fit a parabola to the peak of each CCF to measure the radial velocity. These velocities and their estimated uncertainties are provided in column 4 of Table~\ref{table:rad-vels} under the heading $V_r$(Donor), and these are plotted in Figure~\ref{fig:donor-rv-curve}. We find that the radial velocities follow the expected motion of the mass donor (maximum positive redshift around $\phi = 0.25$, and a maximum negative blueshift around $\phi = 0.75$).

The radial velocities for the donor were then used in the \textit{sbcm} orbit fitting code described by  \citet{Morbey1974} to make a restricted orbital fit.  We assigned equal weights to all the measurements. 
We assumed a circular orbit and set the epoch and period using the light curve (Section~\ref{sec:contemporary-ephemeris}). The derived parameters are a systemic velocity of $\gamma = -27.2 \pm 4.2$ km~s$^{-1}$, an orbital semiamplitude $K_d = 125.8 \pm 6.9$ km~s$^{-1}$, and rms of fit $= 13.4$ km~s$^{-1}$. The residuals (O-C) of this fit are provided in the final column of Table~\ref{table:rad-vels}. Allowing the epoch and eccentricity to vary did not improve the results. The model fit is shown together with the observed radial velocities in Figure~\ref{fig:donor-rv-curve}. The observed scatter is consistent with expectations given the faintness and width of the lines we are measuring.

We expect the mass donor's absorption lines will be broadened due to synchronous rotation with the orbit, so a determination of the donor star's projected rotational velocity $V \sin i$ is an important constraint on the system mass ratio. 
We made fits of rotationally broadened model spectral lines to the observed donor star spectral lines to estimate $V \sin i$. 
The model line profiles were formed by convolving a zero-rotation MARCS model profile with a rotational broadening function for an adopted limb darkening coefficient of $\epsilon = 0.68$ \citep{Wade1985}.  We made small adjustments to the flux scale of the model profile in order to match the local continuum level and central core depth in the observed profile.  Trial fits were made over a grid of $V \sin i$ values, and the case with the minimum $\chi^2$ residuals was selected as the best fit.  
The models were evaluated with spectral lines of the donor star derived from the Doppler tomography reconstructed spectrum of the donor component (discussed in the following section). 
The reconstructed donor spectrum appears in the middle part of Figure \ref{fig:donor-model-lines}. It has better S/N than the single eclipse-phase spectrum, and it better represents the phase-averaged appearance of the donor spectrum.  
We made fits of seven individual line profiles: \ion{Fe}{1} $\lambda\lambda 5473$, 5487, 5569, 5572, 5615, 5624 and \ion{Ca}{1} $\lambda 5594$.
The mean and standard deviation of the fits are $V \sin i = 50.2 \pm 5.5$ km~s$^{-1}$. 
The rotationally broadened model spectrum for this estimate is shown as the lower plot in Figure \ref{fig:donor-model-lines}. 
We will use this value of $V \sin i$ in Section~\ref{sec:discussion} to estimate the binary mass ratio.

A comparison of the line depths of the observed eclipse spectrum with the broadened model spectrum revealed that the donor lines are weaker than expected due to the extra flux within the system that acts to reduce the contrast of the line depths. Assuming that the model $T_{\rm eff}$ is accurate, then a simple depth scaling indicates that the donor spectrum is diluted by an additional continuum flux at $5600$ \AA ~that is approximately $1.9$ times larger than that of the donor. This dilution is more apparent in spectra obtained at out-of-eclipse phases. This additional flux probably originates in the circumstellar disk surrounding the mass gainer.

\subsection{Doppler Tomography Search for Gainer Star Lines}
\label{sec:doppler-tomography-gainer}
The spectral lines of the mass gainer star have eluded observers in prior spectroscopic studies. This is most likely due to the presence of an optically thick accretion torus that surrounds the gainer and blocks the gainer's flux from our view. Nevertheless, the torus itself may present its own unique spectral features that move with the same orbital velocity curve of the underlying gainer star. For example, we might expect the presence of double-peaked emission lines with a central absorption component that are a common disk signature. These features may be relatively weak. Therefore, in order to search for spectral lines associated with gas in the vicinity of the gainer, we used a Doppler tomography algorithm to reconstruct the spectral component that displays the orbital Doppler shifts of the gainer \citep{Bagnuolo1994}. The main spectral components of this system are the shell lines (Section~\ref{sec:shell-lines}), the mass donor's lines (Section~\ref{sec:donor-star-photospheric-lines}), and the lines produced by the mass gainer's circumstellar disk. The radial velocities of the first two components were described above, but we also need to estimate the radial velocity curve of the mass gainer. We can make an initial estimate by assuming that the motion of the mass gainer is opposite to that of the mass donor with a semiamplitude that depends on the mass ratio. We argue in Section~\ref{sec:discussion-mass-ratio-est} that the mass ratio can be estimated from the donor star's projected rotational velocity ($V~\sin~i$) and orbital semiamplitude ($K$) by assuming that the mass donor fills its Roche lobe and rotates synchronously with the orbit. This results in a mass ratio $q = M({\rm donor})/M({\rm gainer}) = 0.36 \pm 0.09$. Thus, the mass gainer's radial velocity curve is the mirror image of the donor's velocity curve with a semiamplitude $K({\rm gainer}) = q K({\rm donor}) = 44.8$ km~s$^{-1}$.  We used this assumed velocity curve to run the tomography algorithm to reconstruct all three spectral components: shell, donor, and gainer.

The observed spectra are mainly the sum of the stellar and torus components attenuated by the CBD shell absorption. However for this application, we make the simplifying assumption that the spectrum is represented by the sum of the shell, donor, and gainer components. In order to avoid instances of strong shell lines reaching near zero flux, we added $1$ to the continuum of the observed spectra prior to running the algorithm. Half of the total flux in this representation is assigned to the shell component while each stellar component is assigned one quarter of the flux. This flux allocation is arbitrary and will only influence apparent line depths in the reconstructed spectra. 
For example, if the actual monochromatic flux ratio $r = f({\rm donor})/f({\rm gainer})$ differs from one, 
then the line depths in the donor (gainer) will be decreased (increased) by a factor of $r$.

\begin{figure*} 
    \centering
    \includegraphics[angle=0,width=15cm]{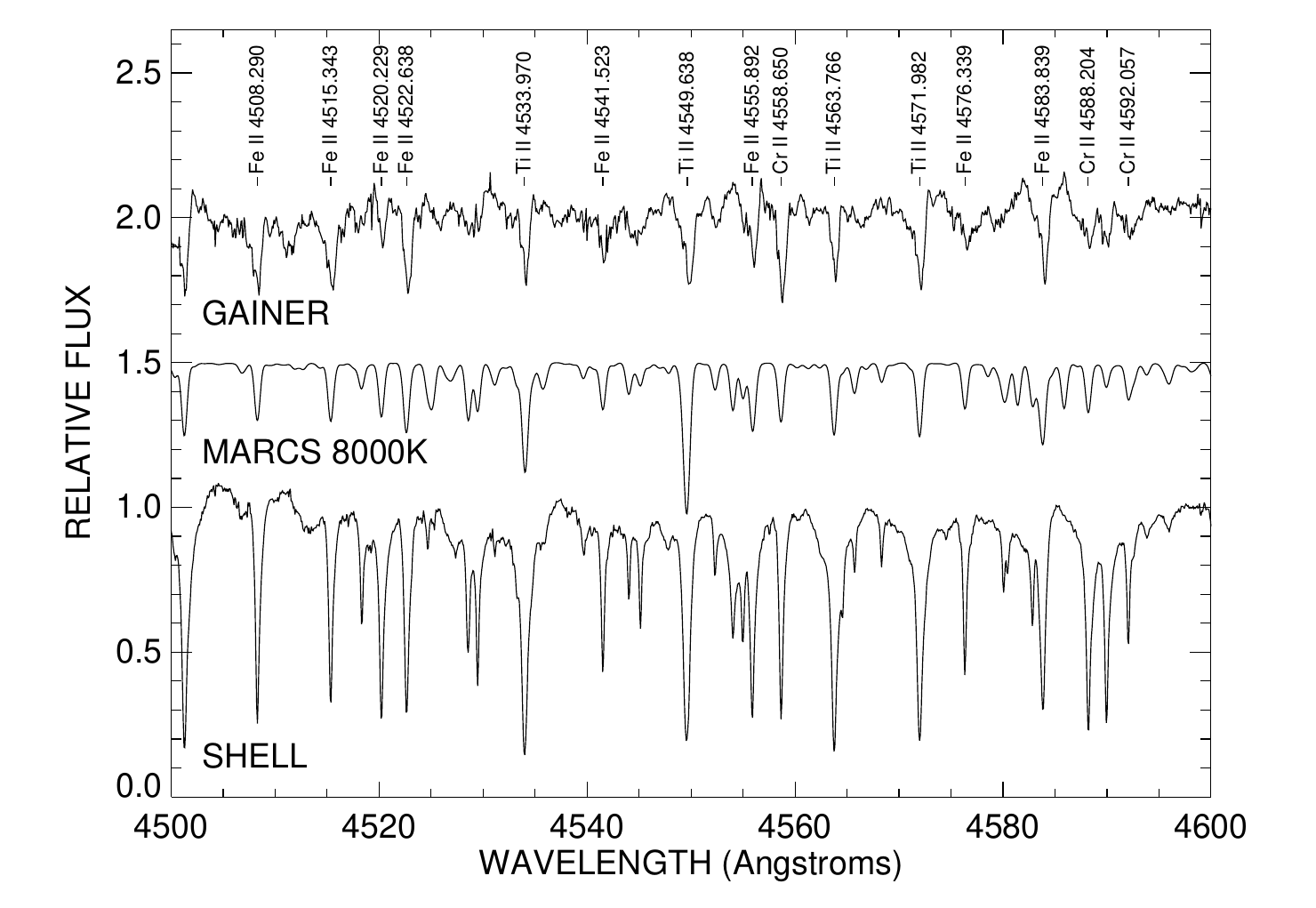}
    \caption{A section of the tomographic reconstructed spectra of the shell component (bottom) and the gainer torus component (top, offset by $+1$ for clarity) along with line identifications of the most prominent lines. An AMBRE/MARCS model spectrum for $T_{\rm eff} = 8000$~K is included between these (offset by $+0.5$ and Gaussian smoothed with FWHM = 40 km~s$^{-1}$). The absorption features for the gainer torus are similar to those of a stellar spectrum that is hotter than that for the circumbinary disk component. }
    \label{fig:plottomabs}
\end{figure*}

\begin{figure*} 
    \centering
    \includegraphics[width=12cm]{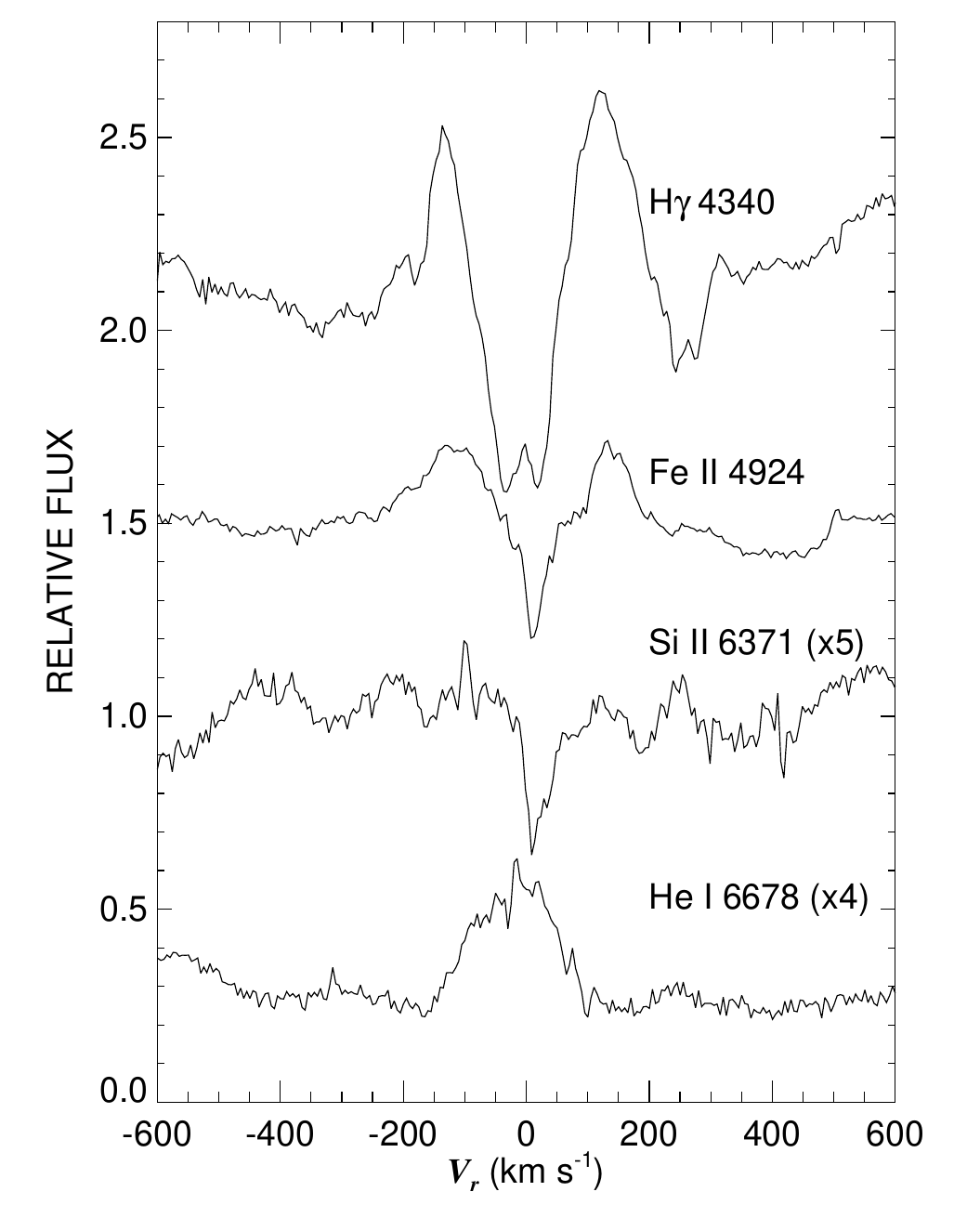}
    \caption{Several features in the reconstructed spectrum of the mass gainer star, with two appearing disk-like. They are offset for clarity by +1.2, +0.5, 0, and $-0.7$ for H$\gamma$ $\lambda 4340$, \ion{Fe}{2} $\lambda 4924$, \ion{Si}{2} $\lambda 6371$, and \ion{He}{1} $\lambda 6678$, respectively. The latter two features appear weak in the reconstructed spectra, so their normalized fluxes are enlarged by factors of 5 and 4, respectively. These features probably form in the torus surrounding the gainer star. }
    \label{fig:plottomem}
\end{figure*}

We then assigned a radial velocity curve to each of these components: the measured CCF velocities for the shell component, the orbital fit velocities for the donor, and the adopted anti-phase velocities for the gainer using the assumed mass ratio. We ran the tomography algorithm for $100$ iterations and a gain of $0.9$ although the convergence of the solution is insensitive to these choices. The reconstructed shell spectrum appears the same as observed in the individual spectra (Fig.~\ref{fig:plottomabs}). The reconstructed donor star spectrum appears weak and noisy, but does show the same absorption lines that are visible in the eclipse spectrum (Fig.~\ref{fig:donor-model-lines}). The reconstructed gainer star spectrum is also weak, but there are a number of features that do appear to be related to its circumstellar torus (Fig.~\ref{fig:plottomabs}, \ref{fig:plottomem}).

A number of narrow absorption lines are found in the reconstructed spectrum for the mass gainer that may form in a pseudo-photosphere of the optically thick circumstellar torus surrounding the gainer (Fig.~\ref{fig:plottomabs}). Also shown in Figure~\ref{fig:plottomabs} is an AMBRE/MARCS model spectrum for $T_{\rm eff} = 8000$~K, $\log g =2.5$, and solar metallicity that was rebinned to the resolution and wavelength grid of the ARCES spectra. There are a number of lines in common in all three spectra presented in Figure~\ref{fig:plottomabs}, however the relative line strengths differ. The model lines make a good match to the reconstructed gainer spectrum. The absorption lines found in the reconstructed gainer spectrum appear similar to those for a $T_{\rm eff} = 8000$~K star, while the CBD shell lines resemble those for a $T_{\rm eff} = 6750$~K star (Section~\ref{sec:shell-lines}). This is consistent with the idea that the features associated with the mass gainer form in the disk surrounding it, where the gas is heated to a hotter temperature than the gas in the circumbinary disk. 

\citet{Ak2007} found a number of lines in the spectrum of the mass gainer in the W~Ser system $\beta$~Lyr that appear disk-like: central absorption bordered by emission wings. We find a number of similar disk-related features in the reconstructed spectrum of the mass gainer. Figure~\ref{fig:plottomem} shows a representative sample of these. H$\gamma$ $\lambda 4340$ shows a deep central absorption with asymmetric emission wings. In individual spectra, this central absorption forms extended absorption wings at $\phi=0.25$ (blue wing) and $\phi=0.75$ (red wing) (see left panel of Fig.\ \ref{fig:hgamma-shellCCF-greyscales}). Similar features are seen in H$\delta$ $\lambda 4101$, \ion{Fe}{2} $\lambda 4924$, and $5018$, although the \ion{Fe}{2} lines have smaller emission wings. H$\beta$ $\lambda 4861$ and H$\alpha$ $\lambda 6563$ have a much more complex appearance in the reconstructed gainer spectrum, presumably due to orbital phase-related variations that are unrelated to Keplerian motion (Section \ref{sec:discussion-visual-model}). The \ion{Si}{2} $\lambda 6371$ and $6347$ lines are both weak but show a trace of emission wings similar to those seen in the stronger lines. \citet{Ak2007} found these \ion{Si}{2} lines to be associated with the torus surrounding the gainer in $\beta$~Lyr. 

Unlike the H-Balmer, \ion{Fe}{2}, and \ion{Si}{2} features that have a morphology indicative of a disk, the \ion{He}{1} $\lambda 6678$ and $\lambda 5876$ features appear to be single-peaked in the reconstructed spectrum. This emission may originate from hot gas that is closer to the mass gainer and may be located in the inner part of the accretion torus surrounding the mass-gainer. In this case we may be seeing the inner edge of the torus on the opposite side of the star. At this location the local gas motion is primarily tangential to our line of sight, creating a single emission peak. Some of these features are visible in both the individual spectra and the reconstructed gainer spectrum (for example, H$\gamma$ $\lambda 4340$ and \ion{He}{1} $\lambda 5876$). These features are among the strongest examples of spectral lines that form in the accretion torus surrounding the mass gainer.


\section{CHARA Array Interferometry}
\label{sec:chara-interferometry}
Despite the long observational history of W~Ser, no interferometric studies have been published to date. This is unfortunate, because interferometry offer us the best chance of resolving either the CBD or the inner binary itself. The successful resolution of either component would allow us to improve greatly upon the current estimates of the system's parameters and further our understanding of its rapid stage of evolution.

The Georgia State University Center for High Angular Resolution Astronomy (CHARA) Array is the world's largest optical/near-IR interferometer with baselines ranging from  $34$ m to $331$ m \citep{tenBrummelaar2005}. It is composed of six $1$-meter telescopes arranged in a $Y$-shaped configuration \citep{tenBrummelaar2005,Schaefer2020} that provides $15$ possible baselines and up to $10$ possible closure phases. Our sample of interferometric observations consists of four nights: 2021 July 28 ($\phi=0.18$), 2023 July 26 ($\phi=0.53$), 2023 July 27 ($\phi=0.60$), and 2023 July 28 ($\phi=0.67$). We used the MIRC-X beam combiner ($H$-band) in all four observations plus the MYSTIC combiner ($K$-band) for the 2023 observations.
The MIRC-X beam combiner \citep{Anugu2020} combines the light from up to six telescopes across $8$ spectral channels (in the prism $R=50$ spectral mode). The MYSTIC beam combiner \citep{Monnier2018,Setterholm2023} combines the light from up to six telescopes at low spectral resolving power (the prism $R=49$ spectral mode). MIRC-X and MYSTIC are capable of operating simultaneously, allowing for increased sensitivity to different portions of the near-IR spectrum. A set of calibrator stars (Table~\ref{table:chara-calibrators}) was also observed each night. 

Both the MIRC-X and MYSTIC data were reduced with the standard pipeline\footnote{\url{https://gitlab.chara.gsu.edu/lebouquj/mircx\_pipeline.git}} \citep{Anugu2020} (version $1.4.0$) using the reduction manual written by J.-B.\ Le Bouquin and C.\ L.\ Davies\footnote{ \url{https://docs.google.com/document/d/1zenNelkhVGTlm1v1tFRvtnb8i0EghIAUeX9F3t5asYU/edit\#heading=h.tnf1bt8jyiiy}}. The pipeline produces OIFITS files that contain both the squared visibilities ($V^{2}$) and the closure phases (CP). The reduced data are then calibrated using the interactive IDL calibration routine created by J.\ D.\ Monnier. This utilizes specified calibrator star observations to estimate the instrumental transfer function and apply it to correct the reduced data. Additionally, we calibrated our calibrators using one another to check for binarity. We performed this check for every night in addition to searching the JMMC Bad Calibrators catalogue\footnote{\url{https://www.jmmc.fr/badcal/}}. Our calibrators were selected using JMMC's Search Cal tool\footnote{ \url{http://www.jmmc.fr/search-cal}} and the parameters used for calibration were obtained from \citet{Bourges2017}. Table~\ref{table:chara-calibrators} lists the date of observation, the HD name of the calibrator, its spectral classification, right ascension (RA) and declination (DEC), Johnson $H$-band magnitude, Johnson $K$-band magnitude, the uniform disk diameter in $H$-band (UD($H$)) and in $K$-band (UD($K$)), and the uncertainty in the uniform disk diameters ($\sigma$(UDD)). 

\placetable{tab2}      
\begin{deluxetable*}{cccccccccc}
\tabletypesize{\scriptsize}
\tablenum{2}
\tablecaption{Calibrator Parameters}
\label{table:chara-calibrators}
\tablewidth{0pt}
\tablehead{
\colhead{Date} & 
\colhead{Name} & 
\colhead{Spectral} &
\colhead{RA} &
\colhead{DEC} &
\colhead{$H$} &
\colhead{$K$} &
\colhead{UD($H$)} &
\colhead{UD($K$)} &
\colhead{$\sigma$(UDD)} \\
\colhead{(UT)} & 
\colhead{} & 
\colhead{Class.} & 
\colhead{(HH:MM:SS)} &
\colhead{(DD:MM:SS)} &
\colhead{(mag)} &
\colhead{(mag)} &
\colhead{(mas)} &
\colhead{(mas)} &
\colhead{(mas)}
}
\startdata
%
2021 Jul 28   & HD 180876   & M3 III  & 19:18:44   & -12:39:37 & 4.38    & 4.16    & 1.012   & 1.035   & 0.088 \\
2023 Jul 26   & HD 189462   & K3 III  & 20:00:36   & -15:52:56 & 5.87    & 5.67    & 0.385   & 0.387   & 0.010 \\
2023 Jul 27   & HD 189462   & \nodata & \nodata    & \nodata   & \nodata & \nodata & \nodata & \nodata & \nodata \\
2023 Jul 28   & HD 177238   & K3 III  & 19:04:20   & -12:33:44 & 5.41    & 5.13    & 0.514   & 0.517   & 0.014 \\
\enddata
\end{deluxetable*}

Our observed $V^2$ and CP measurements are plotted as a function of spatial frequency in Appendix A (Fig.~\ref{fig:vis-t3-pos-gridsearch-all}). For most of our observations of W~Ser, the visibilities are unresolved, $V^{2} \approx 1$, on all baselines except for a small apparent decrease in the longest east-west baselines. Our best night of data is 2023 July 28, when our calculated orbital phase corresponds to an approach towards maximum angular separation. On this night the $V^{2}$ values obviously decrease to below $1$ for most baselines and especially for the longer east-west baselines. This indicates that the observations are beginning to resolve structure on an angular scale similar to the predicted binary separation. 

\subsection{Binary Fits}
\label{sec:chara-gridsearch}
We began by running a basic grid search, binary fit for each night of interferometric data from both MIRC-X and MYSTIC. The binary modeling code {\it gridsearch} was developed by Dr.\ Gail Schaefer\footnote{\url{https://www.chara.gsu.edu/analysis-software/binary-grid-search}} and is described by \citet{Schaefer2016}. It performs an adaptive grid search for a range of separation in RA and DEC, both specified by the user, and at each point within this grid performs a Levenberg-Marquardt least-squares minimization using the IDL routine \textit{mpfit}\footnote{\url{http://cow.physics.wisc.edu/$^\sim$craigm/idl/idl.html}} to calculate the $\chi^{2}$ for the model generated by that position in the grid. These $\chi^{2}$ values are then compared and the RA and DEC positions that produced the lowest $\chi^{2}$ are passed to the user as the result. The program will generally find two best fit locations in the grid, offset by $180^{\circ}$ on either side of the origin. This ambiguity is dependent on whether the brighter or fainter star is adopted as the primary component of the binary. We will assume that the brighter component placed at the origin is the gainer star and its circumstellar torus. Several plots that illustrate these fits are included in Appendix A (Fig.~\ref{fig:vis-t3-pos-gridsearch-all}).

Although the binary is not fully resolved in any of our data, it is helpful to make simple binary fits in order to see if a change in the orbital separation is a viable explanation for the changes between nights in our data. There are nine parameters that define the fit: the spatial offset in RA and in DEC, the flux contributions from each star and from any over-resolved background flux (all summing to unity), the angular diameters of each star, and their limb darkening coefficients. The observations of W~Ser only partially resolve the source, so it is unrealistic to solve for all these parameters. Instead we decided to fix all the parameters except the spatial offsets to a default set of parameters in order to make a highly constrained solution of the binary position only. We then tested how changes to the default set of parameters affect the derived binary position.  

The ratio of the flux contributions of the two components in the near-IR is poorly known. We found in Section~\ref{sec:donor-star-photospheric-lines} that the weak appearance of the donor star's absorption lines indicated that there is about $\approx 1.9$ times more flux in the $V$-band than from the donor star alone. However, this estimate was made for an eclipse-phase observation when other flux sources are partially occulted. Furthermore, the estimate is based upon a comparison to a model with an assumed temperature, and the depths of the model lines may not be applicable if the donor's temperature differs from our estimate. Consequently, there is a large uncertainty surrounding the donor's flux contribution in the near-IR. We make the simplifying assumptions that the extra flux implicated by the donor line depths can be assigned to the gainer and its torus, that the continuum contribution of the CBD is minimal, and that the $V$-band component flux ratio of $f2/f1=f({\rm donor})/f({\rm gainer}) =1/2$ applies across the visible and near-IR bands.

Initial fits of the MIRC-X and MYSTIC $V^2$ data indicate that $V^2(0)<1$ at zero spatial frequency, meaning that there is a component of flux that is over-resolved (generally larger than $\approx 20$ mas). The average value of third light flux that best fits the decreased $V^{2}$ at zero spatial frequency is $f3=0.041 \pm 0.019$. We include this as part of the default parameter set, and using the component flux ratio given above, the other flux components were assigned $f1=0.64$ and $f2=0.32$ for the gainer and donor stars, respectively.  

The final parameters to establish are the angular diameters of the components. Since the binary separation is only partially resolved, we expect the smaller components, such as the gainer, its surrounding torus, and the donor, to be unresolved. Nevertheless, their finite angular size does influence the final results. We used the projected torus and donor areas from the {\it ELC} model (Section~\ref{sec:discussion-prelim-light-curve}) and the Gaia EDR3 distance (Table~\ref{table:system-parameters}) to find equivalent angular diameters of $\theta (1) = \theta ({\rm gainer}) =0.143$~mas and $\theta (2) =\theta ({\rm donor}) =0.155$~mas.
We assumed that both objects are uniform disks for simplicity. 

The binary separation and position angle fits resulting from the default set of parameters are listed in the top part of Table~\ref{table:gridsearch-results}. During the fitting process, we divided the wavelengths in the OIFITS files by systematic correction factors of 1.0054$\pm$0.0006 for MIRC-X and $1.0067 \pm 0.0007$ for MYSTIC (J.~D. Monnier, priv. comm.). We note that the separations listed in Table~\ref{table:gridsearch-results} are below the formal resolution limits of both MIRC-X and MYSTIC (0.5 and 0.6 mas, respectively). We obtained reasonable fits for all of the MIRC-X data, but the derived separation is significantly larger than the uncertainty for only the final night, 2023 July 28. The MYSTIC results were unresolved except on the final night, and the derived separation and position angle are consistent with those from MIRC-X. We next varied the parameters in the default set to determine how the binary positions change for different adopted values. These tests were performed for the 2023 July 28 MIRC-X data, and the results are listed for comparison in the lower rows of Table~\ref{table:gridsearch-results}. The test results show that the fits to the observed decline in the $V^{2}$ curves are sensitive to the adopted parameters. We found that the separation is dependent on three parameters: it increases as the flux ratio $f2/f1$ decreases, as the angular diameters decrease, or as the third light decreases. Thus, the derived binary separations are somewhat dependent on the choices for the parameter set. 

We also explored fits for uniform disk models, uniform elliptical disk models, elliptical Gaussian spatial distributions, and simple image reconstructions (see Appendix B, Fig.\ \ref{fig:macim}).  The results all indicate that the system is partially resolved and non-spherical as expected for an unresolved binary. 
For example, the size and orientation of the elongated ellipse models are consistent with the binary separation and position angle. 

\placetable{tab3}      
\begin{deluxetable*}{ccccccccccc}
\tabletypesize{\scriptsize}
\tablenum{3}
\tablecaption{Interferometric Binary Fits \label{tab3}}
\tablewidth{0pt}
\tablehead{
\colhead{Date} & 
\colhead{Orbital} &
\colhead{Beam} &
\colhead{f1} & 
\colhead{f2} & 
\colhead{f3} & 
\colhead{$\theta (1)$} &
\colhead{$\theta (2)$} &
\colhead{Separation} &
\colhead{PA} & 
\colhead{$\chi_\nu^2$}
\\
\colhead{(HJD-2,400,000)} & 
\colhead{Phase} & 
\colhead{Comb.} & 
\colhead{} & 
\colhead{} & 
\colhead{} & 
\colhead{(mas)} &
\colhead{(mas)} &
\colhead{(mas)} &
\colhead{(deg)} & 
\colhead{}
}
\startdata
%
59423.720 & $0.1817$ & MIRC-X & $0.64$ & $0.32$ & $0.04$ & $0.143$ & $0.155$ & $0.356 \pm 0.005$ & \phn$91.2 \pm 1.7$ & $0.575$ \\
60151.731 & $0.5272$ & MIRC-X & $0.64$ & $0.32$ & $0.04$ & $0.143$ & $0.155$ & $0.164 \pm 0.012$ &  $294.5 \pm 6.4$   & $1.833$ \\
60152.697 & $0.5964$ & MIRC-X & $0.64$ & $0.32$ & $0.04$ & $0.143$ & $0.155$ & $0.192 \pm 0.009$ &  $296.7 \pm 3.7$   & $3.027$ \\ 
60153.698 & $0.6661$ & MIRC-X & $0.64$ & $0.32$ & $0.04$ & $0.143$ & $0.155$ & $0.261 \pm 0.004$ &  $276.1 \pm 1.9$   & $3.058$ \\ 
\vspace{0.2cm}
60153.698 & $0.6661$ & MYSTIC & $0.64$ & $0.32$ & $0.04$ & $0.143$ & $0.155$ & $0.248 \pm 0.007$ &  $272.9 \pm 4.3$   & $0.955$ \\ 
\hline
\vspace{0.2cm}
60153.698 & $0.6661$ & MIRC-X & {\bf 0.80} & {\bf 0.16} & $0.04$ & $0.143$ & $0.155$ & $0.327 \pm 0.006$ & $276.7 \pm 2.1$ & $3.093$ \\ 
\vspace{0.2cm}
60153.698 & $0.6661$ & MIRC-X & $0.64$ & $0.32$ & $0.04$ & {\bf 0.000} & {\bf 0.000} & $0.272 \pm 0.004$ & $274.6 \pm 1.8$ & $3.402$ \\ 
60153.698 & $0.6661$ & MIRC-X & {\bf 0.67} & {\bf 0.33} & {\bf 0.00} & $0.143$ & $0.155$ & $0.317 \pm 0.006$ & $281.8 \pm 1.8$ & $8.189$ \\ 
\enddata
\tablecomments{All parameters are fixed except for the separation and position angle. Formal errors are given here, but more realistic uncertainty ellipses from a bootstrap method are shown in Figure \ref{fig:test-orbit}. The parameters shown in boldface in the final three rows are varied from the default values to show how they influence the results.}
\label{table:gridsearch-results}
\end{deluxetable*}


\subsection{Angular Orbit}
\label{sec:chara-relation-angular-orbit}

\begin{figure*} 
    \centering
    \includegraphics[width=14cm]{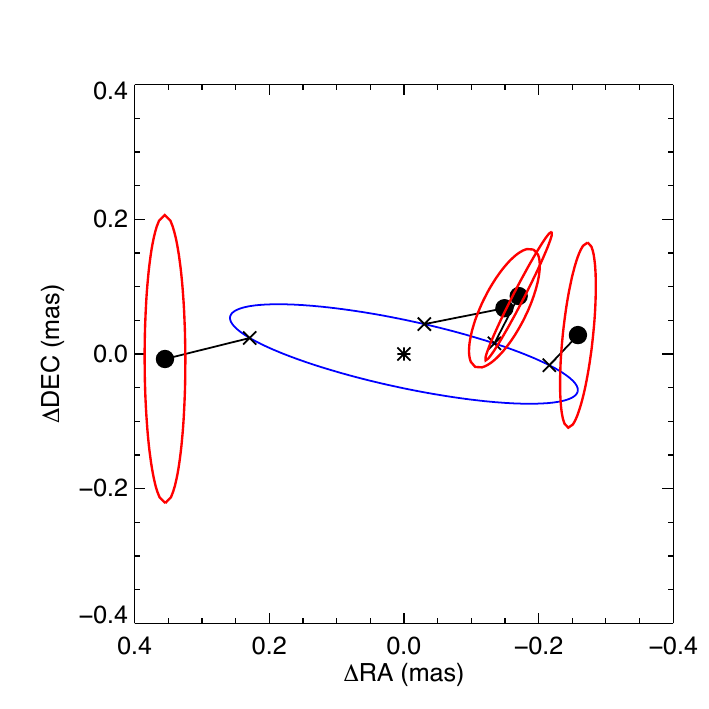}
    \caption{Estimated angular positions of the donor (fainter) relative to the gainer (placed at the origin) from partially resolved interferometry with the CHARA Array. The black circles show the measured positions from fits of the visibility and closure phase (Table \ref{tab3}). The surrounding red ellipses show the nominal error ellipses. The observations are connected to X-symbols that show the predicted positions along the orbit (solid blue) determined from the spectroscopic and light curve analyses and a fit for the longitude of the ascending node $\Omega$. }
    \label{fig:test-orbit}
\end{figure*}

The separations derived from the binary fits are all much smaller than expected for the size of the circumbinary disk, and variations in separation occur on orbital timescales.  Here we show that the implied positions are generally consistent with the predicted binary motion.
Section~\ref{sec:discussion} discusses our analysis of the ASAS light curve (Section \ref{sec:discussion-prelim-light-curve}) and our mass ratio estimate (Section \ref{sec:discussion-mass-ratio-est}). This allows us to predict the angular size (dependent on the masses, period, and distance) and projection (dependent on inclination) of the orbit of the binary. The predicted values are an angular semimajor axis of $a" = 0.264$ mas and an inclination of $i=79\fdg1$ (Table~\ref{table:system-parameters}). We can use these results to obtain a preliminary angular orbit for W~Ser to compare with the interferometric results (Table~\ref{tab3}). This comparison requires determining the sense of motion and the orientation of the orbit in the sky. The 2023 MIRC-X measurements indicate a progressive decrease in position angle with time. This indicates a clockwise orbital motion that is associated with an inclination $>90^\circ$ or $i({\rm CHARA})=180^\circ - i({\rm ELC}) = 100\fdg9$. The orientation of the orbit is set by the longitude of the ascending node $\Omega$. We used the IDL code \textit{newt\_raph\_ell.pro}\footnote{\url{https://www.chara.gsu.edu/analysis-software/orbfit-lib}} described by \citet{Schaefer2016} to fit the MIRC-X positions by varying only $\Omega$ with all of the other orbital elements fixed. The derived result is $\Omega = 78^\circ \pm 12^\circ$, and the predicted orbit is compared to the CHARA measurements in Figure~\ref{fig:test-orbit}. The uncertainties of the fitted positions were computed through a Monte Carlo bootstrap technique in which a sample of visibility and closure phase measurements were selected at random with repetition and varied within their uncertainties. A binary model was fit to the resampled measurements, and this process was repeated 1000 times. 
The error ellipses plotted in Figure~9 show the major and minor axes and position angles computed for the 67\% confidence interval of the bootstrap distribution. We find that the CHARA measurements appear to be consistent with the predicted orbit, although we caution that our measurements are sensitive to the adopted parameters used in the fits (Table~\ref{tab3}).  
The measurements appear to indicate an orbit that is slightly larger than predicted.  There are several possible explanations 
including the assumed distance and choice of fitting parameters.  It is also possible that the center of light measured by interferometry is displaced from the geometric position of the gainer to the vicinity of the outer L3 region that is a source of flux in some emission lines (Section 5.5).  It will be valuable to repeat the interferometric measurements in concert with spectroscopic and light curve observations that are sensitive to mass loss to determine how 
much the spatial light distribution is influenced by gas flows.

\newpage
\section{Discussion}
\label{sec:discussion}

\subsection{Summary of the System Components} 
\label{sec:discussion-spectral-components}
Here we review the observational results presented above that support the presence of three main flux components in the binary: the accretion torus around the gainer, the photosphere of the donor star, and an extended circumbinary disk. The component with the intrinsically brightest flux is probably the hot mass gainer star, however most of its flux is blocked by a surrounding optically thick accretion torus. The torus itself is probably the dominant flux source across most of the visible and near-infrared spectrum, and a tomographic reconstruction of spectral features that share the orbital motion of the gainer shows evidence of absorption lines that are formed in the pseudo-photosphere and emission lines formed in hotter regions of the torus (Section~\ref{sec:doppler-tomography-gainer}). Additionally, we identified for the first time the photospheric lines of the mass donor (Section~\ref{sec:donor-star-photospheric-lines}). The line strengths of the donor vary with orbital phase, and the lines are most visible during the primary eclipse. The associated velocity curve (Fig.\ \ref{fig:donor-rv-curve}) indicates that the mass donor is the occulting object at primary eclipse. The donor is probably a cool star that fills its Roche lobe and is actively transferring mass to the region surrounding the mass gainer. The circumbinary disk (CBD) is responsible for the strong shell lines seen in the spectrum. The shell lines exhibit no orbital motion (Fig.~\ref{fig:hgamma-shellCCF-greyscales} and Table~\ref{table:rad-vels}) because the Keplerian motions of the CBD are mainly orthogonal to our line of sight and the resulting Doppler shifts are small. Below we discuss the derivation of the component mass ratio (Section~\ref{sec:discussion-mass-ratio-est}) and the orbital inclination (Section~\ref{sec:discussion-prelim-light-curve}) that we use to determine the masses and other parameters of the binary (Section~\ref{sec:discussion-sys-dimensions}). We then present a conceptual model of the mass transfer and mass loss from the binary (Section~\ref{sec:discussion-visual-model}), and we discuss the spectroscopic evidence supporting the existence of mass outflow.  

\subsection{Mass Ratio Estimation} 
\label{sec:discussion-mass-ratio-est}
Although the velocity curve for the gainer is unavailable, we can estimate the mass ratio indirectly \citep{Gies1986}. If we assume that the donor is filling its Roche lobe, then the donor will be tidally locked with its companion and will rotate synchronously with the orbit. The projected rotational velocity of the donor is then
\begin{equation} \label{eq:4}
    V_{d} \sin i = \omega R_{d} \sin i = \omega r_{L} a \sin i
\end{equation}
where $\omega$ is the orbital frequency, $\omega = 2 \pi / P$, $R_{d}$ is the average radius of the donor, $r_{L}$ is the fractional Roche lobe radius that depends on the mass ratio $q=M({\rm donor})/M({\rm gainer})$ (from \citealt{Eggleton1983}), $a$ is the binary semimajor axis, and $i$ is the orbital and rotational axis inclination. If we compare equation \ref{eq:4} to the measured semiamplitude of the donor's orbital motion, $K_d = (\omega a \sin i)/(1 + q)$, then their ratio is independent of inclination and depends only on the mass ratio: 
\begin{equation}
    \frac{V_d \sin i}{K_d} = (1 + q) r_{L}(q).
\end{equation}
Using $V_{d} \sin i$ and $K_{d}$ from Section~\ref{sec:donor-star-photospheric-lines}, we derive a mass ratio $q=0.36 \pm 0.09$. This indicates that the donor has transferred most of its mass to the mass-gainer. This mass ratio inversion from $q > 1$ to $q < 1$ is consistent with an increasing orbital period \citep{Erdem2014}. We used this value of $q$ in Section~\ref{sec:doppler-tomography-gainer} to estimate the velocity curve of the gainer and make a tomographic reconstruction of spectral features associated with the motion of the gainer.

\subsection{Preliminary Light Curve Analysis} 
\label{sec:discussion-prelim-light-curve}
In order to determine individual masses from the mass ratio and the mass function (from the radial velocity curve of the donor), we need to estimate the orbital inclination of the system. We expect $i \approx 80^\circ$ from the extreme depth of the primary eclipse \citep{Guinan1989}. We made preliminary fits of the $V$-band light curve from ASAS (Fig.\ \ref{fig:asas-light-curve-elc-fit}) using the {\it ELC} code described by \citet{Orosz2000}. We limited these fits by setting all but a few parameters in the light curve model. The parameters we fixed include $P$, $T$, $K_{d}$, $q$, $e=0$, and $T({\rm donor})=5000$ K. We additionally assumed that the donor fills its Roche lobe, and the radiative intensity has a Planck curve distribution. 

For our first trial, we assumed that the gainer is a simple star with a typical B-type star temperature of $T({\rm gainer})=17900$ K, and that there is no third light contributing to the flux. This leaves two parameters available for fitting: the fill-out factor for the gainer that is needed to match the eclipse duration (and determine gainer radius $R_{g}$), and the orbital inclination $i$ that is varied to match the eclipse depth. Our results for this fit are $R_{g} = 13.2~R_{\odot}$ and $i=78\fdg0$. This fit is over-plotted as the solid line in Figure \ref{fig:asas-light-curve-elc-fit}. The secondary eclipse, which occurs when the donor is eclipsed by the gainer, is shallow due to the relatively low temperature of the donor and its low surface intensity. 

For our second trial, we noted that the large size of the gainer produced by the first fit is unrealistic for a main sequence B-type star. We therefore reduced the size of the gainer to a fixed $R_g = 3.8~R_{\odot}$ and introduced an optically thick disk surrounding the gainer. This accretion torus is implemented in {\it ELC} as a flared disk with a temperature that decreases with distance from the gainer \citep{Orosz1997}. The disk blocks flux from all sources behind it, and adds flux according to the projection of the disk elements in the sky, their temperature, and orientation to the line of sight. In general, this is seen as the unocculted disk edge in front of the gainer star, and the opposite face of the disk projected behind the star. The disk extends from the surface of the gainer $R[torus]({\rm inner})=R_{g}$ to $R[torus]({\rm outer})$, which is predicted to be approximately half of the gainer's Roche radius \citep{Lu2023}. The disk temperature is set to vary as $\propto (r/R[torus]({\rm inner}))^{-0.425}$ from $14300$~K at $R[torus]({\rm inner})$ to $8000$~K at $R[torus]({\rm outer})$. The outer disk temperature is set to the temperature associated with the model spectrum that matched the shell features in the reconstructed spectrum of the gainer and its torus (Section~\ref{sec:doppler-tomography-gainer}). The opening angle of the upper disk rim was set to $\beta_{\rm rim} = 25^\circ$. This parameter primarily controls the portion of the photosphere of the gainer 
that is blocked from view by the disk. The gainer is totally obscured from view for this disk flare angle and the derived orbital inclination.

We are again left with two fitting parameters: $R[torus]({\rm outer})$ which sets the duration of the eclipse of the disk by the donor, and $i$ which determines the eclipse depth. Our results for this second fit are $i=79\fdg1$, $R[torus]({\rm outer}) / (a~r_{L}(1/q)) = 0.65$, and $R[torus]({\rm outer}) = 14.9 R_{\odot}$. This model is over-plotted as the thick gray line in Figure~\ref{fig:asas-light-curve-elc-fit}. Despite the different kinds of flux sources being occulted by the donor at primary eclipse, both light curve models give reasonable fits and both are associated with similar inclination angles.

\subsection{System Dimensions} 
\label{sec:discussion-sys-dimensions}
We can now utilize the orbital mass function, the mass ratio, and the inclination from the disk model of the light curve to obtain the individual masses of the binary. These masses and other system parameters are presented in Table~\ref{table:system-parameters}. The uncertainties for a number of the results are not listed due to the inherent dependence on certain assumptions that are difficult to quantify (such as the Roche lobe filling and synchronous rotation of the donor). Table~\ref{table:system-parameters} lists the following: the orbital elements of period, time of primary eclipse, the systemic velocity $\gamma$ and semiamplitude of the donor's orbital motion, eccentricity, mass function $\mathcal{F}(M)=(M_{g}^{3} \sin^{3} i)/(M_{d}+M_{g})^{2}$, the mass ratio obtained from the $(V_d \sin i)/(K_d)$ ratio, the inclination obtained from the {\it ELC} disk model of the ASAS light curve, the semimajor axis, the masses, radii, adopted effective temperatures of the stars, the mass gainer's circumstellar disk temperature range and radial and vertical dimensions, the Gaia EDR3 distance \citep{Bailer-Jones2021}, and the estimated angular semimajor axis associated with this distance. The final two rows give the longitude of the ascending node $\Omega$ and the orbital inclination for clockwise motion in the sky that we derived from CHARA Array interferometric observations. 

\placetable{tab4}      
\begin{deluxetable*}{ccc}
\tabletypesize{\scriptsize}
\tablenum{4}
\tablecaption{System Parameters}
\label{table:system-parameters}
\tablewidth{0pt}
\tablehead{
\colhead{Parameter} & 
\colhead{Unit} & 
\colhead{Value}
}
\startdata
%
 $P$        & days        & $14.1787$       \\
 $T$        & HJD-2,400,000 & $60002.678$     \\
 $\gamma$   & km s$^{-1}$ & $-27 \pm 4$     \\
 $K_d$      & km s$^{-1}$ & $126 \pm 7$     \\
 $e$        & \nodata     & $0$             \\
 $\mathcal{F}(M)$& $M_\odot$   & $2.9 \pm 0.5$   \\
 $q$        & $M_d / M_g$ & $0.36 \pm 0.09$ \\
 $i$(ELC)   & deg         & $79.1$          \\
 $a$        & $R_\odot$   & $48.7$          \\
 $M_d$      & $M_\odot$   & $2.0$           \\
 $M_g$      & $M_\odot$   & $5.7$           \\
 $R_d$      & $R_\odot$   & $14.3$          \\
 $R_g$      & $R_\odot$   & $3.8$           \\
 $T_d$      & kK          & $5$             \\
 $T_g$      & kK          & $17.9$            \\
 $T[torus]$(outer) & kK          & $8$             \\
 $T[torus]$(inner) & kK          & $14.3$          \\
 $R[torus]$(outer) & $R_\odot$   & $14.9$          \\
 $R[torus]$(inner) & $R_\odot$   & $3.8$           \\
 $z[torus]$(outer) & $R_\odot$   & $6.9$           \\
 $z[torus]$(inner) & $R_\odot$   & $1.8$           \\
 $d$        & pc          & $857$           \\
 $a"$       & mas         & $0.264$         \\
 $\Omega$   & deg         & $78 \pm 12$      \\ 
 $i$(CHARA) & deg         & 100.9           \\
\enddata
\end{deluxetable*}

\begin{figure*} 
    \centering
    \includegraphics[width=15cm]{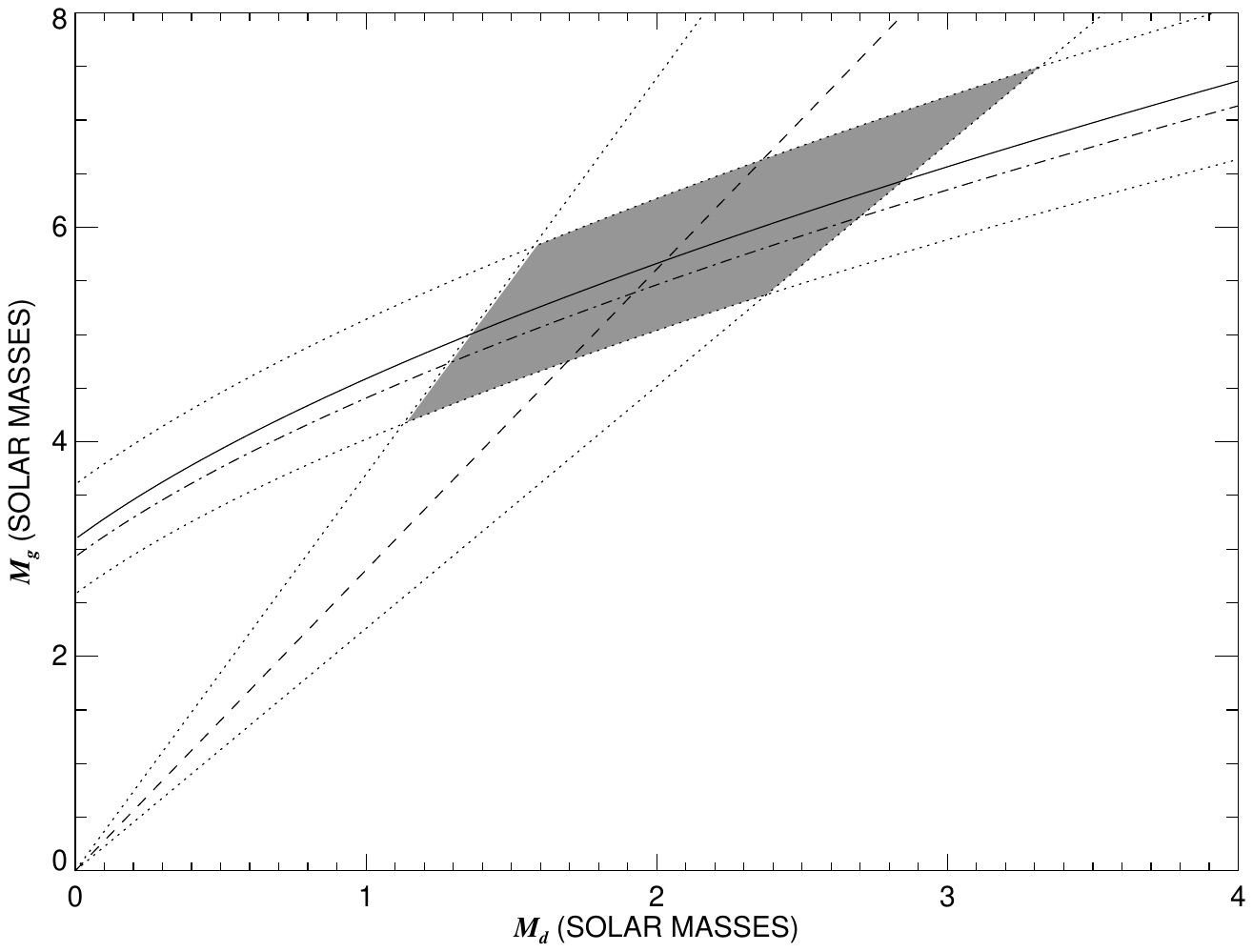}
    \caption{Mass of the donor and the gainer based upon the donor star spectroscopic mass function $\mathcal{F}(M)$. The curved lines correspond to the best fit inclination of $i=79\fdg1$ (solid line) and the lower limit for $i=90^\circ$ (dash-dot line). The dotted lines represent the range given the uncertainty in $\mathcal{F}(M)$. The dashed diagonal line indicates a mass ratio of $q=0.36 \pm 0.09$ (dotted lines for uncertainty). The intersection point of the solid and dashed lines marks the best fit, and the gray shaded region indicates the error limited range of solutions. }
    \label{fig:mass-plot}
\end{figure*}

Figure~\ref{fig:mass-plot} shows the relationship between the gainer and donor mass based upon the the derived mass function $\mathcal{F}(M)$ and its uncertainty. The solid line tracks the relationship for the inclination from the light curve analysis, $i=79\fdg1$, and the dot-dashed line shows that for the minimum mass case of $i=90^\circ$. The dashed diagonal line shows the mass ratio derived from the $(V_d \sin i) / K_d$ ratio. The best estimate of the masses occurs at the intersection of the solid and dashed lines at a donor mass of $M_{d}=2.0~M_{\odot}$ and a gainer mass of $M_{g}=5.7~M_{\odot}$. These estimates are generally larger than have been quoted in past studies \citep{Erdem2014,Mennickent2016,Davidge2023}. The mass of the hidden gainer is comparable to that of an early B-type star, which supports our assumption regarding the star's hot temperature. 

The CHARA Array interferometric measurements appear to confirm the expectations about the angular orbit from the semimajor axis, inclination, and distance as summarized in Table~\ref{table:system-parameters} (Fig.~\ref{fig:test-orbit}). Because the binary is not fully resolved in the CHARA $H$-band observations, the uncertainty in the angular semimajor axis is substantial and depends upon assumptions made about the flux ratio, background flux, and component angular sizes. However, the interferometry does provide first estimates about the orientation of the orbit in the sky and evidence for a small flux component in the near-infrared from an over-resolved source (possibly the CBD). 

\subsection{Mass Loss Model for W Ser} 
\label{sec:discussion-visual-model}
\begin{figure*} 
    \centering
    \includegraphics[width=15cm]{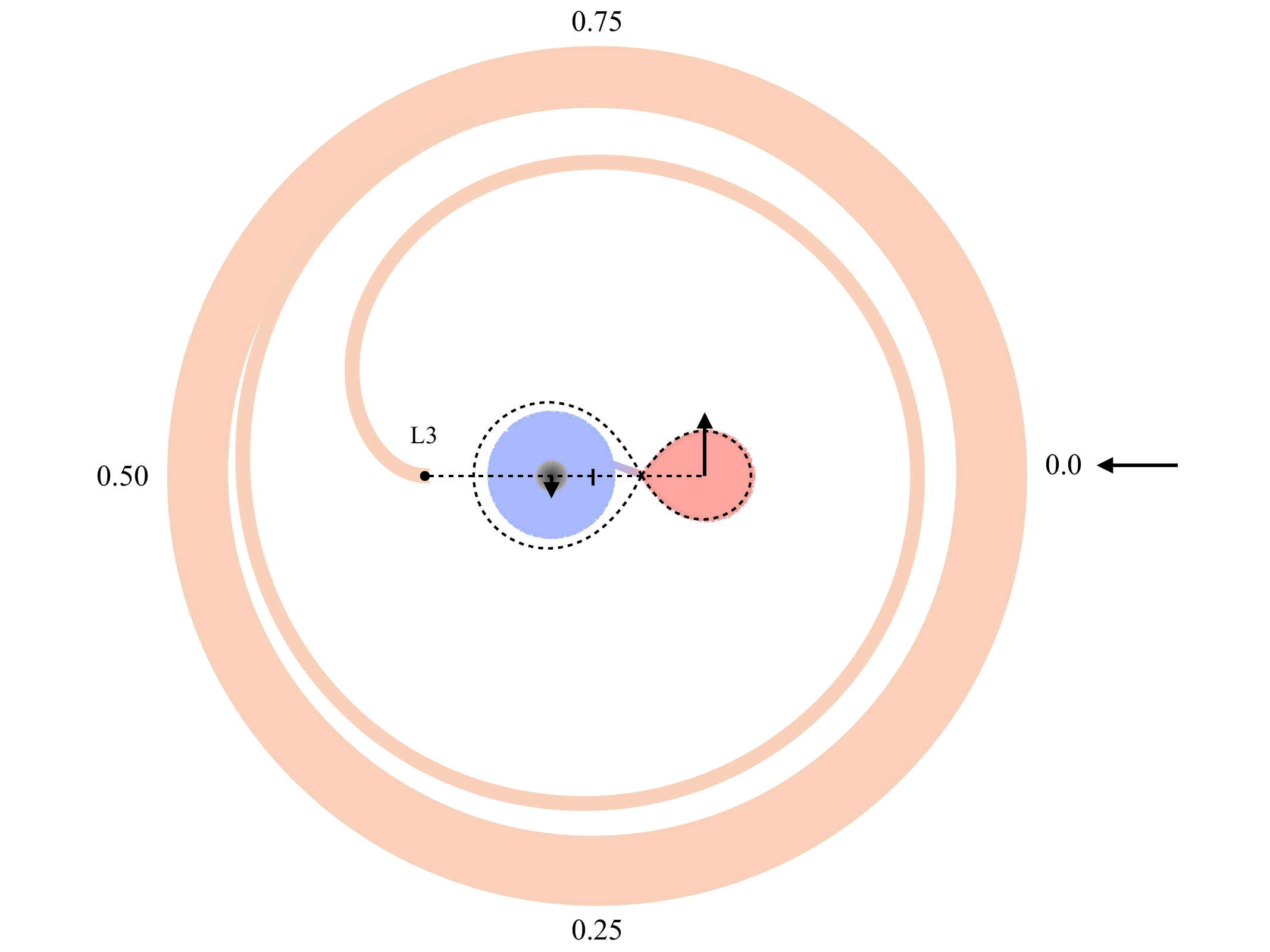}
    \caption{A graphical representation of the configuration of W~Ser as seen from above. The mass donor is on the right and is filling its Roche lobe. The mass gainer (gray) is on the left and is surrounded by its accretion torus (light 
    blue). The mass transfer stream lies between the stars, 
    but its trajectory and point of impact with the torus may be time variable and related to the stochastic variability of the light curve.
    The center of mass is indicated by a tick mark. Material leaking out of the region near the mass gainer forms a spiral structure starting at the L3 point. This spiral gradually progresses outwards until it joins the circumbinary disk. The size of the CBD relative to the inner binary is not to scale. Numbers on the periphery refer to the orbital phases for different lines of sight.}
    \label{fig:my-model}
\end{figure*}

The results outlined above rely on the assumption that the donor star fills its Roche lobe. This is a secure assumption given the presence of circumstellar gas as evidenced in the strong emission line spectrum, the lack of gainer star spectral features due to a surrounding torus, the deep shell lines formed in a CBD,
and the observed period increase.
The picture that emerges is sketched out in Figure~\ref{fig:my-model}. This portrayal depicts the approximate system dimensions as viewed from above the orbital plane. The donor (right) fills its Roche lobe and overflowing gas forms a mass transfer stream towards the gainer (left). This gas stream encounters a large and thick torus surrounding the gainer. The gainer itself is similar to a main sequence star and is much smaller than its Roche lobe. We expect that the gainer has been spun up by previous mass transfer to near critical rotation making it difficult for the gainer to accrete any additional gas from the surrounding torus. Consequently, the torus acts as a holding zone for accreted gas until some of the material leaks out through the outer L3 Lagrangian point on the opposite (left) side of the gainer. We see this outflow best when it is in the foreground around $\phi=0.5$. Any gas lost through this Lagrangian point gradually spirals outwards and merges into the CBD. The projection of the CBD against the components of the binary produces the main shell spectrum that is observed. 

Hydrodynamical simulations of gas flows in Roche-filling binaries offer some insights about the mass transfer and mass loss processes. For example, \citet{Nazarenko2005} present simulations in two and three dimensions of gas flows in massive Algol-type binaries, and these account for gas cooling processes in systems of differing separation and gainer radius. The general result is the presence of an outflowing and trailing gas stream that emerges in the vicinity of L3. Their example of a simulation of the RZ~Sct system (see their Fig.~3) may be relevant for the case of W~Ser. In this model, the mass flow from the donor to the gainer is deflected by the dense torus around the gainer and into an axial outflow through L3. The gas is expected to be hot where the gas stream strikes the accretion disk, while on the opposite side of the gainer near the L3 point, the gas is cooler, denser, and has an outwards velocity. This gas outflow forms a spiral structure that follows behind the mass gainer and progresses outwards into the CBD.

There are a number of spectral features that provide clues about the L3 mass stream. Consider, for example, the very strong H$\alpha$ emission line shown in Figure~\ref{fig:greyscales-halpha-HeI5876}. The H$\alpha$ profile displays excess emission in the redshifted wing around $\phi=0.0$. Half an orbit later at $\phi=0.5$, we see the opposite, a slight excess emission in the blueshifted wing. Both of these extended wings appear to reach a velocity of approximately $200$ km~s$^{-1}$. This Doppler shift is similar to that observed in the \ion{Si}{4} $\lambda 1400$ feature at the same phases by \citet{Weiland1995}. The \ion{O}{1} $\lambda 7775$ line (Fig.~\ref{fig:greyscales-CaII8498-OI7775}) shows similar redshifted emission at $\phi=0.0$, but instead shows extended blueshifted absorption at $\phi=0.5$ (suggesting gas projected against a brighter background). These features are best explained as the result of an extended gas outflow near L3 that appears redshifted at $\phi=0.0$ and blueshifted at $\phi=0.5$ (see Fig.~\ref{fig:my-model}). The gas stream crossing the gap between the stars has a similarly directed flow velocity, but we doubt that this contributes significantly to the observed profiles at conjunctions because that region is occulted by the donor and torus at $\phi=0.0$ and $\phi=0.5$, respectively.  

We also see evidence of a greater velocity dispersion in features formed near L3 that may reflect increased gas pressure and turbulence in this vicinity. For example, we observe that the individual absorption components of the \ion{O}{1} $\lambda 7775$ feature become much more blended in spectra obtained around $\phi=0.5$ (Fig.~\ref{fig:greyscales-CaII8498-OI7775}). Broadening of the absorption core around $\phi=0.5$ is also observed in the upper members of the Paschen series, such as P11 $\lambda 8862$. We expect that most of the shell line formation occurs in the CBD, but it is possible that cooler gas in the vicinity of L3 may also contribute. Recall that we observed the development of a weak but extended red wing in the CCFs constructed from the shell lines that occurs after $\phi=0.5$ (Fig.~\ref{fig:hgamma-shellCCF-greyscales}). This variation was much more apparent 80 years ago when \citet{Bauer1945} obtained spectra of W~Ser. We suggest that some of the cool gas near L3 that moves with the gainer is projected against the torus to create shell features that share some of the motion of the gainer after the conjunction at $\phi=0.5$. 
Alternatively, both the orbital and long-term variations in the kinematics of the circumbinary disk lines may be the result of the development of asymmetries in the inner part of the circumbinary disk that are shown to form in numerical simulations of disks surrounding binaries \citep{Penzlin2022}.

We note for completeness that the Na D doublet also shows evidence of axial outflows in both directions at the conjunction phases (see the weak absorption components in the line wings shown in the right panel of Fig. \ref{fig:greyscales-halpha-HeI5876}).  This may indicate there is also some mass loss occurring near the L2 location beyond the far side of the donor star. 

The gas lost through the L3 region will accumulate in a large and extended circumbinary disk (CBD)
with a vertical extension driven by thermal expansion \citep{Lu2023}. 
The spectrum displays many disk-like but stationary features that are probably formed in the large CBD. The H$\alpha$, \ion{Ca}{2}, and \ion{Fe}{2} lines all display stationary double peaked emission. These lines form at different temperatures that probably correspond to the changing disk temperature over a range in radius (cooler at larger radius). We see this reflected in the separations of their emission peaks. The emission lines with smaller peak separation, which tend to be weaker lines, form further out in the disk where the gas is cooler (fainter) and moving slower. The outer regions of the CBD may be excellent environments for the production of dust that can emit in the mid- and far-infrared \citep{Geisel1970}. In fact, \citet{Davidge2023} found mid-IR dust emission in the WISE W2 images of W~Ser that extends over 20 arcseconds from the star. The CBD represents the net gas lost and funneled out through the L3 zone during the current phase of non-conservative mass transfer and mass loss.

\newpage
\section{Conclusions}
\label{sec:conclusions}
W~Serpentis is a key example of how interacting binary stars are transformed during a short but critical stage of intense mass transfer. The new spectroscopic and interferometric observations presented here have led to the first detection of the mass donor star. This has allowed us to make estimates of the previously unknown system parameters. We find that the deepest absorption (shell) lines in the visible spectrum are formed in a dense circumbinary disk. These lines show small radial velocity variations that are unrelated to orbital motion. We have discovered a set of weak absorption lines that are associated with a cool temperature $\approx 5000$ K, whereas the strong shell lines are associated with a temperature closer to $\approx 6750$ K. These weak lines show the orbital velocity variations expected for the cool, mass donor that eclipses the mass gainer at primary eclipse, $\phi=0.0$. The donor star spectral lines are rotationally broadened, and assuming the donor fills its Roche lobe and is rotating synchronously with the orbit, we use the ratio of its projected rotational velocity to the orbital semiamplitude to derive a mass ratio $M_d / M_g = 0.36 \pm 0.09$.

We also fit the ASAS eclipsing light curve of W~Ser by adopting all of the orbital and physical properties of the components except for the inclination and radius of the gainer. Two trial fits were made by assuming first a large star for the gainer and then a smaller gainer surrounded by a thick accretion disk. Both models yield an inclination of $i \approx 79^{\circ}$ and masses of $M_{d} = 2.0 M_{\odot}$ and $M_{g} = 5.7 M_{\odot}$. The binary orbital elements, mass estimates, and GAIA EDR3 distance predict an angular orbit that appears to be consistent with the partial angular resolution of the binary in CHARA Array interferometric measurements. 

We did not detect any spectral features directly related to the gainer star. However, we used a Doppler tomography algorithm to reconstruct the spectral lines of a component with the predicted Doppler shifts of the gainer. This reconstructed spectrum is marked by shell lines associated with a hotter plasma ($\approx 8000$~K) and by several emission features that have a double-peaked structure that is associated with a disk.  We suggest that these spectral features form in an optically thick accretion torus that surrounds the gainer and effectively blocks it from view. 

Many of the spectral lines show features of blue-shifted outflow around $\phi=0.5$ when the mass gainer is in the foreground. We interpret these Doppler shifts as evidence of systemic mass loss in the vicinity of the L3 Lagrangian point that feeds gas into a much larger CBD. Dust formation in the outskirts of the CBD forms a mid-infrared halo that is detected in WISE observations \citep{Davidge2023}. This indicates that much of the mass lost by the donor escapes from the binary instead of being accreted by the gainer.

W~Ser is experiencing an important stage of systemic mass loss in its binary evolution. As the donor is further stripped of its envelope, its tidal influence will wane, and large scale mass loss through the L3 location may shut down \citep{Lu2023}. Depending on the outcome of this mass transfer stage, we expect that the system will become a longer period binary consisting of a rapidly rotating B-type star and a hot companion stripped of its envelope. Thus, W~Ser is a probable progenitor of the Be + sdO binaries that have recently been detected through FUV spectroscopy \citep{Wang2021} and CHARA interferometry \citep{Klement2024}.

\begin{acknowledgments}
This work is based upon observations obtained with the Georgia State University Center for High Angular Resolution Astronomy Array at Mount Wilson Observatory. The CHARA Array is supported by the National Science Foundation under Grant No. AST-1636624, AST~1908026, and AST-2034336. Institutional support has been provided from the GSU College of Arts and Sciences and the GSU Office of the Vice President for Research and Economic Development.

This work has made use of data from the European Space Agency (ESA) mission {\it Gaia} (\url{https://www.cosmos.esa.int/gaia}), processed by the {\it Gaia} Data Processing and Analysis Consortium (DPAC, \url{https://www.cosmos.esa.int/web/gaia/dpac/consortium}). Funding for the DPAC has been provided by national institutions, in particular the institutions participating in the {\it Gaia} Multilateral Agreement.

SK acknowledges funding for MIRC-X from the European Research Council (ERC) under the European Union's Horizon 2020 research and innovation programme (Starting Grant No. 639889 and Consolidated Grant No. 101003096). JDM acknowledges funding for the development of MIRC-X (NASA-XRP NNX16AD43G, NSF-AST 1909165) and MYSTIC (NSF-ATI 1506540, NSF-AST 1909165).

This research has made use of the Jean-Marie Mariotti Center \texttt{Aspro} service (available at \url{http://www.jmmc.fr/aspro}), and \texttt{Search Cal} service (available at \url{http://www.jmmc.fr/search-cal}). 

\end{acknowledgments}

\newpage
\appendix
\vspace{-0.9cm}
\section{CHARA Data and Associated Gridsearch fits}
\begin{figure*}[h!]
        \centering
	\includegraphics[height=16cm,trim={5cm 0 5cm 0}]{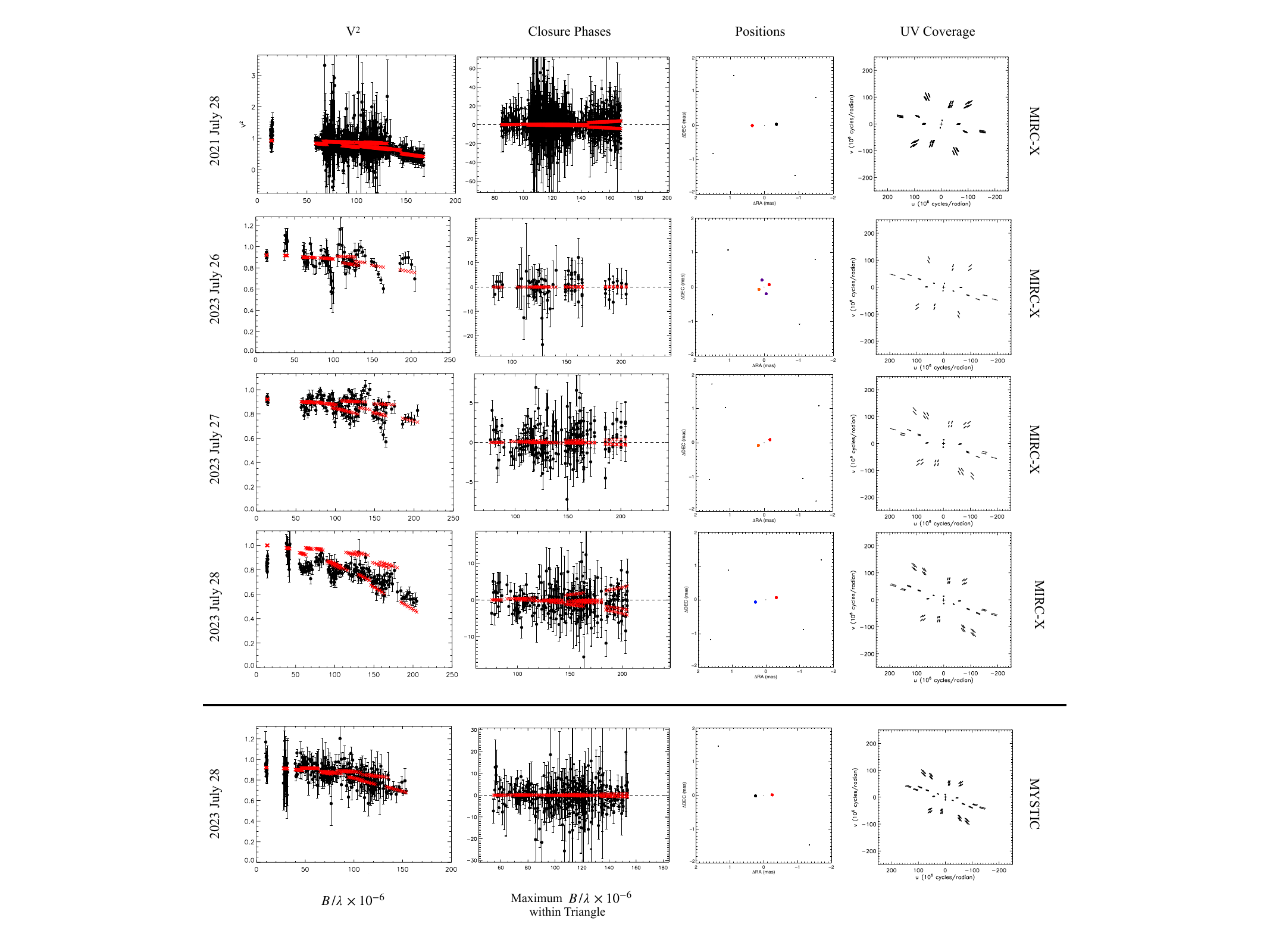}
	\caption{The CHARA Array data for W~Ser are plotted as black points, while the best fit results from {\it gridsearch} are over-plotted in red. The visibilities are plotted in the left column, the closure phases are plotted in the second column, the positions associated with the best fit result from {\it gridsearch} are plotted in the third column (RA on the x-axis, and DEC on the y-axis), and the spatial frequency $(u,v)$ coverage is plotted in the final column. The brighter component (not shown) is located at the origin in the position diagrams. Since we fixed the fluxes of each component, the code generally finds two solutions for the companion that differ by $180^{\circ}$ in position angle. The red circle in the $\chi^2$ maps identifies the solution with the lowest $\chi^2$ and gives the preferred position of the fainter companion relative to the brighter component. The first four rows show data from MIRC-X, and the bottom row shows data from MYSTIC on the best night. From top to bottom, the data were obtained on 2021 July 28, 2023 July 26, 2023 July 27, and 2023 July 28 (for MIRC-X and MYSTIC). Note that there are two pairs of solutions for the 2023 July 26 data set. We selected the better $\chi^2$ solution that was also more consistent with the predicted angular orbit (Fig. \ref{fig:test-orbit}).  
        }
        \label{fig:vis-t3-pos-gridsearch-all}
\end{figure*}

\pagebreak

\section{MACIM Image Reconstructions of CHARA Data}
\begin{figure*}[h!]
        \centering
	\includegraphics[width=\textwidth]{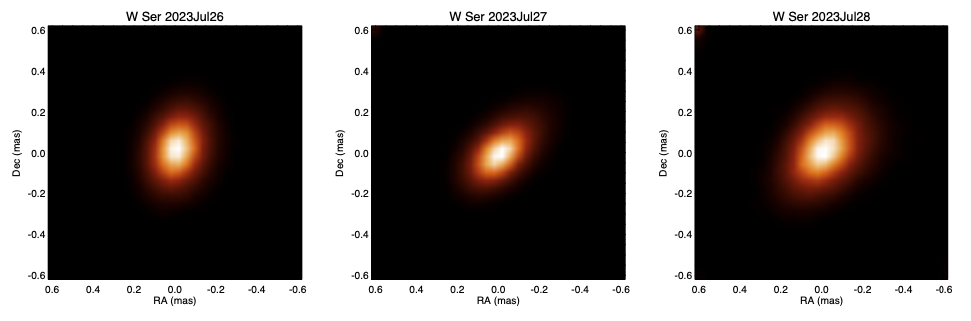}
	\caption{MACIM image reconstructions \citep{Ireland2006} of the three 2023 nights of data for W~Ser made by J.\ D.\ Monnier. Each image has $0.04$ mas per pixel. These images illustrate not only the change in position angle that is related to binary motion, but also how difficult it is to resolve this system.}
        \label{fig:macim}
\end{figure*}

\facilities{CHARA, APO}

\software{AMBRE/MARCS, ELC, IDL, IRAF, MACIM, SBCM}


\bibliography{ms1}{}
\bibliographystyle{aasjournal}

\end{document}